\documentclass[apj]{emulateapj}
\usepackage{natbib}
\usepackage{graphicx}
\usepackage[toc,page]{appendix}
\usepackage{color}
\usepackage{threeparttable}
\usepackage{hyperref}
\usepackage{breakurl}
\usepackage{amsmath}
\begin{document}

\slugcomment{Accepted for publication in the Astrophysical Journal Supplements}
\title{The Hubble Legacy Field GOODS-S Photometric Catalog}
\email{kwhitaker@astro.umass.edu}
\author{Katherine E. Whitaker\altaffilmark{1,2,3}, Mohammad Ashas\altaffilmark{2}, Garth Illingworth\altaffilmark{4}, 
Daniel Magee\altaffilmark{4},  Joel Leja\altaffilmark{5}, Pascal Oesch\altaffilmark{6,3}, Pieter van Dokkum\altaffilmark{7},
Lamiya Mowla\altaffilmark{7}, Rychard Bouwens\altaffilmark{8}, Marijn Franx\altaffilmark{8}, Bradford Holden\altaffilmark{4}, 
Ivo Labb\'{e}\altaffilmark{9},
Marc Rafelski\altaffilmark{10,11}, Harry Teplitz\altaffilmark{12}, Valentino Gonzalez\altaffilmark{13}}
\altaffiltext{1}{Department of Astronomy, University of Massachusetts, Amherst, MA 01003, USA}
\altaffiltext{2}{Department of Physics, University of Connecticut, Storrs, CT 06269, USA}
\altaffiltext{3}{Cosmic Dawn Center (DAWN)}
\altaffiltext{4}{UCO/Lick Observatory, University of California, Santa Cruz, CA 95064, USA}
\altaffiltext{5}{Harvard-Smithsonian Center for Astrophysics, 60 Garden Street, Cambridge, MA 02138, USA}
\altaffiltext{6}{Department of Astronomy, University of Geneva, Ch. des Maillettes 51, 1290 Versoix, Switzerland}
\altaffiltext{7}{Department of Astronomy, Yale University, New Haven, CT 06511, USA}
\altaffiltext{8}{Leiden Observatory, Leiden University, NL-2300 RA Leiden, Netherlands}
\altaffiltext{9}{Swinburne University of Technology, Hawthorn, VIC 3122, Australia}
\altaffiltext{10}{Space Telescope Science Institute, 3700 San Martin Drive, Baltimore, MD 21218, USA}
\altaffiltext{11}{Department of Physics \& Astronomy, Johns Hopkins University, Baltimore, MD 21218, USA}
\altaffiltext{12}{Infrared Processing and Analysis Center, MS 100-22, Caltech, Pasadena, CA 91125, USA}
\altaffiltext{13}{Departmento de Astronomia, Universidad de Chile, Casilla 36-D, Santiago 7591245, Chile}
\shortauthors{Whitaker et al.}

\begin{abstract}
This manuscript describes the public release of the Hubble Legacy Fields
(HLF) project photometric catalog for the extended GOODS-South region from the
Hubble Space Telescope (HST) archival program AR-13252. The analysis is
based on the version 2.0 HLF data release that now includes all ultraviolet (UV)
imaging, combining three major UV surveys. The HLF data combines
over a decade worth of 7475 exposures taken in 2635 orbits totaling 6.3
Msec with the HST Advanced Camera for Surveys Wide Field Channel
(ACS/WFC) and the Wide Field Camera 3 (WFC3) UVIS/IR Channels in the
greater GOODS-S extragalactic field, covering all major observational
efforts (e.g., GOODS, GEMS, CANDELS, ERS, UVUDF and many other programs; see
Illingworth et al 2019, in prep). The HLF GOODS-S catalogs include photometry in
13 bandpasses from the UV (WFC3/UVIS F225W, F275W and F336W
filters), optical (ACS/WFC F435W, F606W, F775W, F814W and F850LP
filters), to near-infrared (WFC3/IR F098M, F105W, F125W, F140W and F160W
filters). Such a data set makes it possible to construct the spectral
energy distributions (SEDs) of objects over a wide wavelength range from
high resolution mosaics that are largely contiguous. Here, we describe a
photometric analysis of 186,474 objects in the HST imaging at
wavelengths 0.2--1.6$\mu$m. We detect objects from an ultra-deep image
combining the PSF-homogenized and noise-equalized F850LP, F125W, F140W
and F160W images, including Gaia astrometric corrections. 
SEDs were determined by carefully taking the effects
of the point-spread function in each observation into account. All of
the data presented herein are available through the HLF website
(https://archive.stsci.edu/prepds/hlf/).
\end{abstract}

\keywords{catalogs --- galaxies: evolution --- galaxies: general --- methods: data analysis --- techniques: photometric}

\section{Introduction}
\label{sec:intro}

Our current understanding of the formation and evolution of galaxies with cosmic time
is driven by large, statistical samples that span a broad range of multi-wavelength
observations.  The deepest and highest resolution observations exploring the peak epoch of 
star formation in our universe are
those from the Hubble Space Telescope \citep[HST; e.g.,][]{Giavalisco04, Scoville07, Grogin11, Koekemoer11, Momcheva15}.  
When combining HST with the deepest ground-based
observations and Spitzer Space Telescope, surveys enable the measurement of fundamental galaxy
properties for tens of thousands of extragalactic sources.  HST alone has pushed
galaxy studies into uncharted territory \citep[e.g.,][]{McLeod15,Oesch16}.

The scientific returns from extragalactic legacy surveys are maximized
when data sets are combined in a homogeneous way. To this end, we
undertake the construction of a photometric catalog based solely on all
high resolution HST imaging taken in the greater GOODS-S extragalactic field to date. 
While the future inclusion of \emph{Spitzer}/IRAC and ground-based ancillary data
will continue to improved the measured photometric redshifts and stellar population parameters, 
this work serves as a necessary albeit incremental step towards a comprehensive final catalog of
the GOODS-S extragalactic legacy field.
The extended GOODS-S/CDF-S region has the largest ensemble of HST imaging
data of any area of the sky.  The equivalent of approximately 75\% of an HST cycle has
now been committed to imaging this area through more than 30 different
programs. In total, there is 6.3 Msec of HST on-target time
through 7475 exposures taken over 2635 orbits of ACS, WFC3/IR and
WFC3/UVIS imaging. A summary of all  programs is found in Table~\ref{tab:programs}.

\begin{figure*}[t]
\leavevmode
\centering
\includegraphics[width=\linewidth]{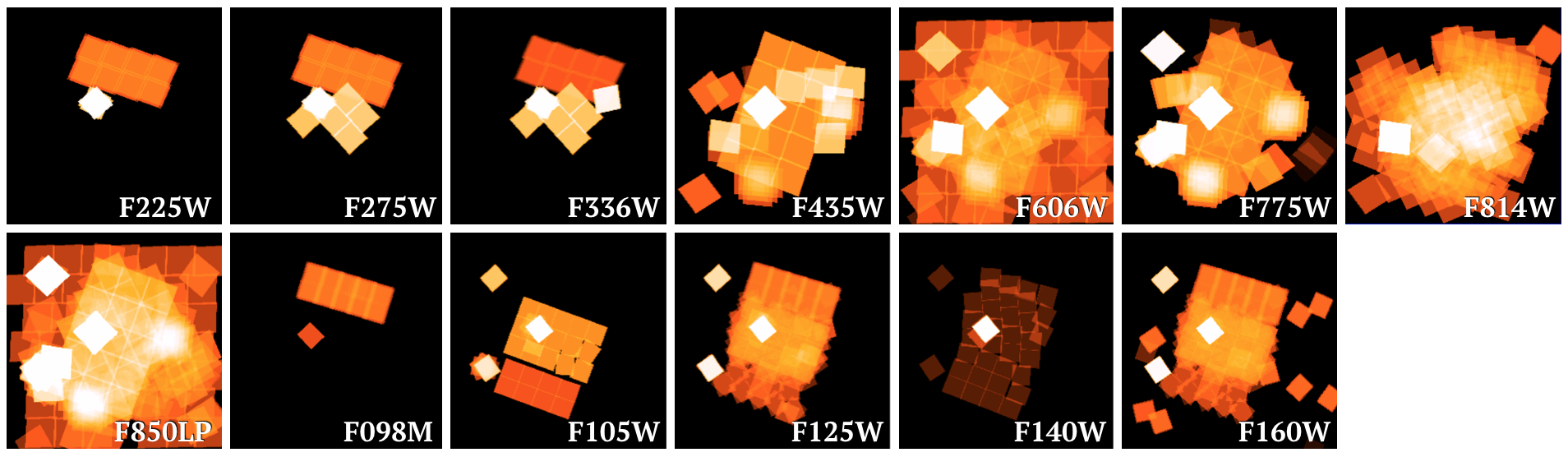}
\caption{The HLF-GOODS-S dataset weight maps outlining the footprints for the three WFC3/UVIS, five ACS/WFC, and
five WFC3/IR filters.  White represents the deepest data corresponding to the footprint of the HUDF/XDF.}
\label{fig:exptime}
\end{figure*}

The catalog is based on the version 2.0 (v2.0) release of the Hubble Legacy Field
GOODS-S (HLF-GOODS-S). Figure~\ref{fig:exptime} shows the coverage for the thirteen
HST filters included in the photometric catalog.  The v2.0 version of HLF-GOODS-S updates the v1.5
version with the inclusion of all of the available UV imaging data. 
The three UV surveys added constitute a
substantial body of data, totaling 213 orbits of HST WFC3/UVIS
imaging, or about 0.5 Msec of observations: the Early
Release Science (ERS) observations \citep{Windhorst11}, the
UltraViolet Ultra-Deep Field (UVUDF) dataset \citep{Teplitz13,Rafelski15}, 
the Hubble Deep UltraViolet (HDUV) legacy dataset \citep{Oesch18}, as well 
as additional F336W imaging data \citep{Vanzella16}.
A summary of these UV programs and the details of all other datasets from v1.5
can be found in Table~\ref{tab:programs}. The orbit values listed 
are computed from the total exposure time in each program/filter, where 1 orbit 
equals 2400s of exposure time. The UV datasets
were updated and astrometrically-matched to the v1.5 release of the
HLF-GOODS-S.  The ERS dataset of \citet{Windhorst11} required a full processing as high level science 
products are not available on the Mulkulsi Archive for Space Telescopes. 
The steps that were taken to assemble the HST UV, optical and near-IR data,
including details of the data reduction and astrometric
analysis, can be found in Illingworth et al. (2016, 2019, in prep).
Here, we describe the details of the source
detection, PSF homogenization, and catalog construction. We provide the
homogenized set of images that are used in this paper to the community,
in addition to the photometric catalog.

The structure of this manuscript is as follows. In Section 2.2 and 2.3, we describe the additional
background subtraction and the source detection, respectively.
Section 2.3 details the PSF matching of the different resolution images, and Section 2.4 the general layout
of the photometric catalogs themselves. We present basic internal and external diagnostic plots to 
verify quality and consistency in Section 3. 
Section 4 contains a general overview of the public release of the HLF GOODS-S photometric catalog.

In this manuscript, all magnitudes are in the AB system and we assume a $\Lambda$CDM 
cosmology with $\Omega_{\mathrm{M}}=0.3$, $\Omega_{\Lambda}=0.7$, and $\mathrm{H_{0}}=70$ km s$^{-1}$ Mpc$^{-1}$.

\begin{table*}[!th] 
\centering                                                                                                                               
\caption{Hubble Space Telescope programs contributing to the HLF-GOODS-South}
\begin{tabular}{llll}                                                                                                                    
\hline \hline                                                                                                                            
\noalign{\smallskip}                                                                                                                     
Program ID & Program & Filter(s) & Orbit(s) \\                                                                                         
\hline                                                                                                                                   
\noalign{\smallskip}  
9352       & \nodata     & F606W/F775W/F850LP                                           & 2/2/12                  \\
9425       & GOODS       & F435W/F606W/F775W/F850LP                                     & 45/33/33/68             \\
9480       & \nodata     & F775W                                                        & 12                      \\
9488       & \nodata     & F775W/F850LP                                                 & 3/2                     \\
9500       & GEMS        & F606W/F850LP                                                 & 56/60                   \\
9575       & \nodata     & F775W                                                        & 3                       \\
9793       & GRAPES      & F606W                                                        & 1                       \\
9803       & HUDF-NICMOS & F435W/F606W/F775W/F850LP                                     & 17/19/35/52             \\
9978       & HUDF        & F435W/F606W/F775W/F850LP                                     & 52/54/139/137           \\
9984       & \nodata     & F775W                                                        & 1                       \\
10086      & HUDF        & F435W/F606W/F775W/F850LP                                     & 4/2/6/8                 \\
10189      & PANS        & F435W/F606W/F775W/F850LP                                     & 1/5/7/17                \\
10258      & \nodata     & F606W/F775W/F850LP                                           & 11/1/24                 \\
10340      & PANS        & F606W/F775W/F850LP                                           & 2/12/48                 \\
10530      & \nodata     & F606W                                                        & 5                       \\
10632      & HUDF-P1/P2  & F606W/F775W/F850LP                                           & 18/45/138               \\
11144      & \nodata     & F125W/F850LP                                                 & 1/1                     \\
11359      & ERS         & F225W/F275W/F336W/F814W/F098M/F125W/F140W/F160W     & 19/19/9/17/21/21/0/21   \\
11563      & HUDF09      & F435W/F606W/F775W/F814W/F850LP/F105W/F125W/F160W    & 18/42/40/30/79/50/77/98 \\
12007      & \nodata     & F606W                                                        & 1                       \\
12060      & CANDELS     & F606W/F814W/F850LP/F105W/F125W/F160W                        & 11/31/14/61/2/2         \\
12061      & CANDELS     & F814W/F850LP/F125W/F160W                                    & 70/9/42/44              \\
12062      & CANDELS     & F606W/F814W/F850LP/F125W/F160W                               & 2/50/12/33/34           \\
12099      & CANDELS-SN  & F435W/F606W/F775W/F814W/F850LP/F098M/F105W/F125W/  & 1/2/1/19/4/1/1/10/1/8   \\
   & & F140W/F160W & \\
12177      & 3D-HST      & F814W/F140W                                                  & 8/13                    \\
12461      & CANDELS-SN  & F125W/F160W/F435W/F606W/F814W/F850LP                         & 4/1/0/1/2/2             \\
12498      & HUDF12      & F105W/F140W/F160W/F814W                                      & 83/34/30/135            \\
12534      & UVUDF       & F225W/F275W/F336W/F435W/F606W/F775W/F814W/F850LP            & 18/17/16/73/5/2/12/2    \\
12866      & \nodata     & F160W/F814W                                                  & 13/11                   \\
12990      & \nodata     & F160W                                                        & 1                       \\
13779      & \nodata     & F105W/F435W/F606W/F814W                                      & 8/4/2/4                 \\
13872      & HDUV       & F275W/F336W/F435W                                            & 50/45/47                \\
14088      & \nodata     & F336W                                                        & 20                                           \\ 
\noalign{\smallskip}                                                                                                                     
\hline                                                                                                                                   
\noalign{\smallskip}                                                                                                                     
\end{tabular}                                                                                                                            
\label{tab:programs}                                                                                                                     
\vspace{+0.5cm}    
\end{table*}

\section{Photometry}
\label{sec:data}

We construct the HLF photometric catalog as detailed below, closely following the techniques discussed
in depth in \citet{Skelton14} and \citet{Shipley18}. 
In summary, we use a deep noise-equalized combination of the four HST bands (F850LP, F125W, F140W, F160W)
for detection. 12 HST bandpasses (F225W, F275W, F336W, F435W, F606W, F775W, F814W, F850LP, F098M, F105W, F125W, and F140W) are 
each convolved to the F160W point-spread function (PSF) in order to
measure consistent colors across all wavebands. For this entire analysis, we use the v2.0 release 60 mas pixel
scale mosaics.
Aperture photometry was performed in dual-image mode
using Source Extractor \citep{Bertin96} on the background-subtracted, homogenized 
images using a small aperture of diameter of 0.7$^{\prime\prime}$ that maximizes the signal-to-noise of the 
resulting aperture photometry.

\subsection{Background Subtraction}

With the v2.0 mosaics for the optical and near-infrared filters (F435W--F160W) and the ultraviolet (F225W--F336W), 
we first do an additional sky subtraction to remove 
any excess light previously missed during the initial routine sky subtraction performed
during the data reduction.  The sky subtraction is performed using 
Source Extractor \citep[SExtractor;][]{Bertin96}, using a Gaussian interpolation of the background 
with an adopted mesh size of 64 pixels and a 7 pixel median filter size. 
The result of this sky subtraction is on the order of a few
hundredths of a percent per pixel, a minimal correction but necessary to improve
the overall homogeneity of the background.

\subsection{Source Detection}
\label{sec:detection}

We create a detection image that is a noise-equalized version of the mosaics combining
one ACS (F850LP) with three WFC3 bands (F125W, F140W, F160W) by multiplying the PSF-matched science images (see
details in Section~\ref{sec:PSF}) by the square root of the inverse variance map. 
These four noise-equalized images are then coadded to form an ultra-deep detection image. 
Such a methodology has been adopted in several earlier surveys: e.g., NMBS \citep{Whitaker11}, 3D-HST \citep{Skelton14}, 
and HFF-DeepSpace \citep{Shipley18}.  
The decision to include F850LP stems from the wide field coverage in this filter that extends to significantly
larger area than the nominal WFC3 footprint.  Our methodology adopts an extremely deep detection image
while also explicitly taking into account variations in the weight across the mosaics.  This variation in weight 
is a natural consequence of combining many different observing programs with unique science goals into single mosaics.
We use a detection and analysis threshold of 1.8$\sigma$ and 1.4$\sigma$, respectively, and require a minimum area of 14 pixels for detection. 
The deblending threshold is set to 32, with a minimum contrast parameter of 0.0001. 
A Gaussian filter of 7 pixels is used to smooth the images before detection. 
The detection parameters were optimized such that the settings are a 
compromise between deblending neighboring objects while minimizing dividing larger objects 
into multiple components. Moreover, visual inspection confirms that the input SExtractor parameters find all faint objects in the
ultra-deep detection images, while limiting the number of spurious detections.

The resulting objects detected are not cleaned for spurious detections within SExtractor itself, as 
this may cause subtle problems with the segmentation maps.  Instead, we clean the photometry in post-processing.
Any object residing in a region with a weight less than 1\% of the 95th percentile weight is identified as problematic 
and the photometry of the respective
band is fixed to a value of -99.  This represents 30\% of the total catalog in the F160W and F850LP filters. 

\begin{figure*}[t]
\leavevmode
\centering
\includegraphics[width=\linewidth]{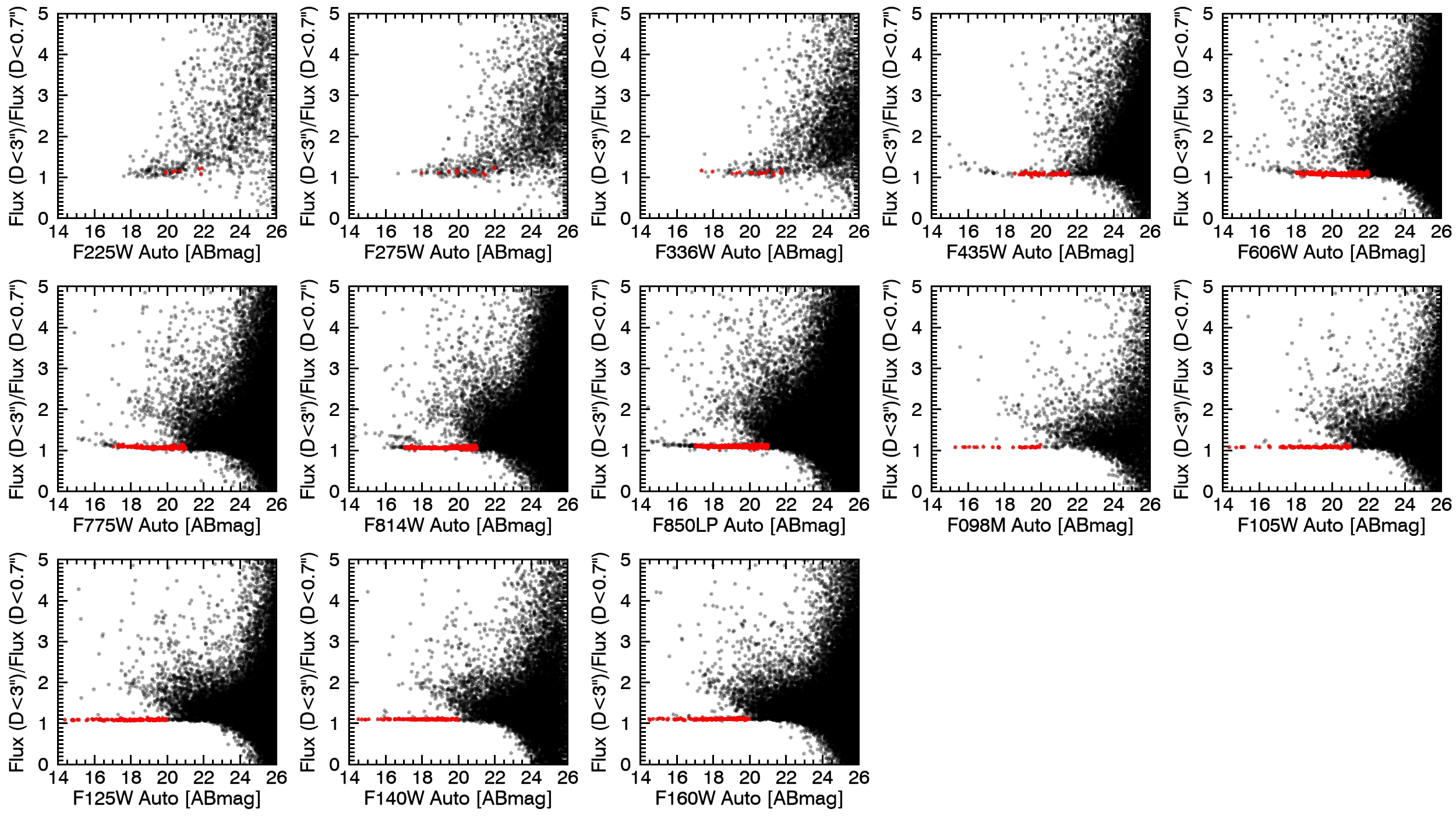}
\caption{Point sources have a ratio of flux within a larger 3$^{\prime\prime}$ diameter
circular aperture relative to a small 0.7$^{\prime\prime}$ diameter aperture close to unity.  
We can therefore cleanly identify bright, unsaturated stars (red) to generate empirical PSFs in each 
filter on the basis of this ratio.}
\label{fig:stars}
\end{figure*}

\begin{figure*}[t]
\leavevmode
\centering
\includegraphics[width=\linewidth]{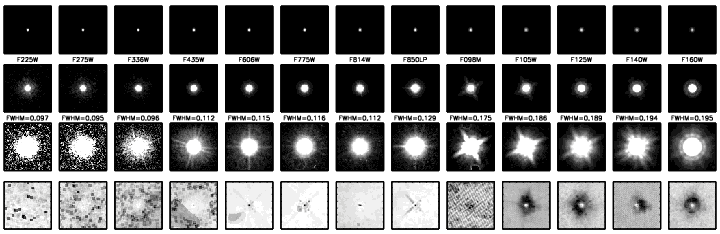}
\caption{(Top) empirical point spread functions derived from bright stars selected across each mosaic, displayed at 
different stretch levels from the top to third row to highlight various features.
(Bottom) stacked weight maps for the stars used to derive the PSF. }
\label{fig:psf}
\end{figure*}

\subsection{Point Spread Function Homogenization}
\label{sec:PSF}

In order to measure accurate colors, we need to PSF-match the HST ACS and WFC3 images to the 
filter with the broadest FWHM (the F160W filter in this dataset) prior  to performing aperture photometry. To do so, empirical PSFs are 
created for each HST image by stacking isolated unsaturated stars selected from across the mosaic.  
An initial sample of stars are identified on the basis of the ratio of their fluxes within a large 3$^{\prime\prime}$ diameter
aperture relative to a small 0.7$^{\prime\prime}$ diameter aperture.  Bright, unsaturated stars 
can be cleanly identified from this ratio (see Figure~\ref{fig:stars}).  The number of stars identified
ranges from a minimum of 29 in F098M to a maximum of 353 in F606W, with a 
more typical value of 100--200 stars. 
We extract 5$^{\prime\prime}$$\times$5$^{\prime\prime}$ regions around each star, recentering and masking
nearby pixels that are either associated with nearby objects according to the segmentation map, 
or 5$\sigma$ above the local noise.  All stars that either require $>$1.5 pixel shifts to recenter or fail altogether are additionally rejected 
at this stage. 
From this parent sample,
we perform a visual inspection to remove any remaining problematic stars.  A few examples of removed sources include cases where the central
pixels were masked incorrectly as cosmic rays, the objects fell on the edge of the detector, or the object was severely contaminated by
nearby bright objects.
Most stars automatically identified in the UVIS F225W--F336W filters fell on the
edges of the detector and therefore failed the earlier recentering algorithm.  For example, after this step, the total number of useable stars reduced 
from 100s to 8--21 in F225W-F336W.
The median local background is measured on the final stacked image for pixels located at a 
radius of 4--5$^{\prime\prime}$.  Though often negligible, we subtract this background correction. 
We do not attempt to take into account variations with chip position, as the mosaics comprise 
multiple pointings with different orientations and overlap.  As noted by \citet{Skelton14}, we
expect such differences to be small.  

Figure~\ref{fig:psf} shows the empirical PSFs and weight maps generated from the procedure outlined herein based
on the v2.0 HLF project mosaics.  Note that there may be some residual false clipping of the central pixels of the ACS
images due to the cosmic ray rejection adopted during the data reduction.  Evidence for this can be
seen in the depression in the centers of the stacked ACS weight maps for bright stars. 
One reason this happens in the ACS images is that the dataset itself has been taken over 
an approximately 12 year timespan. Over this epoch, the positions of some stars have changed. 
The central pixels will therefore get clipped, as they are no longer aligned due to this proper motion.
The top row in Figure~\ref{fig:psf} emphasizes the core of the PSF where most of the power resides, whereas
the contrast in middle and bottom rows highlights the first Airy ring (∼0.5\%) and the diffraction spikes (∼0.1\%), respectively.  
A single orientation would contain four diffraction spikes resulting from the secondary mirror assembly.
We see in some cases here a much larger number of diffraction spikes (especially for WFC3) due to the 
broad range of orientations that comprise the mosaiced data.  The trade off of the aggressive deblending adopted
on the ultra-deep detection image is that the diffraction spikes and first Airy ring around bright stars will
often be identified as a separate object from the star itself.  This can be seen in the weight maps, with 
the masked regions outlining these PSF features.  Note however that these are very faint features and the
PSF remains robust given the large number of stars contributing to the stack in the NIR; the point of the PSF homogenization is
to match the light profiles across all of the filters, which we will show in the next section is good to the $<$0.5\% level at all radii.

\begin{figure}[t]
\leavevmode
\centering
\includegraphics[width=\linewidth]{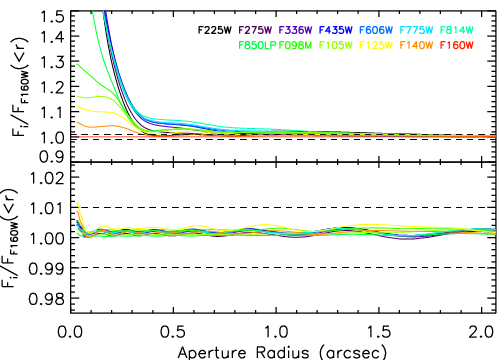}
\caption{Growth curves showing the fraction of light enclosed as a function of 
radius for each HST filter relative to the F160W growth curve before (top) and after (bottom) convolution.}
\label{fig:growthcurve}
\end{figure}

\begin{figure}[t]
\leavevmode
\centering
\includegraphics[width=\linewidth]{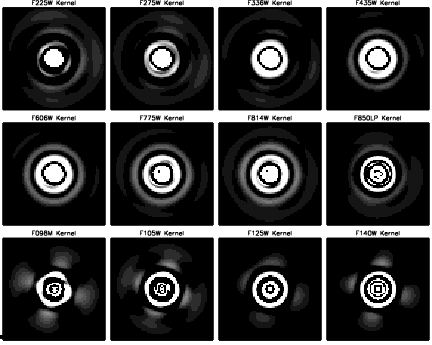}
\caption{Convolution kernels derived using a linear combination of Gaussian-weighted Hermite polynomials
to match each empirical PSF to the broadest FWHM F160W filter.}
\label{fig:kernel}
\end{figure}

Figure~\ref{fig:growthcurve} shows the curve of growth, defined as the fraction of light enclosed 
as a function of aperture size, for each of the PSFs, normalized at 2$^{\prime\prime}$. The top panel
of Figure~\ref{fig:growthcurve} shows the growth curves from the empirical PSFs presented in Figure~\ref{fig:psf},
whereas the bottom panel shows the results after convolving each PSF to match F160W.  We derive the convolution
kernel by fitting a series of Gaussian-weighted Hermite polynomials to the
Fourier transform of the empirical PSFs (Figure~\ref{fig:kernel}).  This methodology yields PSFs with almost indistinguishable 
growth curves on the scales of interest, agreeing to $<$0.5\% at all radii.

\begin{figure}[t]
\leavevmode
\centering
\includegraphics[width=\linewidth]{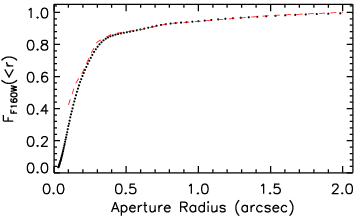}
\caption{The fraction of light enclosed as a function of radius for the F160W PSF,
relative to the total light within 2$^{\prime\prime}$. The red dashed line shows the 
encircled energy as a function of aperture size, also normalized to 2$^{\prime\prime}$, 
from the WFC3 handbook. The empirical growth curves (black points) agree well with the 
theoretical expectation.}
\label{fig:handbook}
\end{figure}

Finally, we present a comparison between the encircled energy as a function of aperture provided in the WFC3
handbook relative to our derived F160W empirical PSF in Figure~\ref{fig:handbook}.  The marginal
deviations towards the center of the PSF are not significant, with the curves showing excellent agreement.

\subsection{Detection Limits}

It is challenging to define completeness limits given our detection methodology 
and the nature of the HLF dataset, combining a wide range of surveys with dramatically 
varying depths and coverage between filters.  Moreover, defining a single-band magnitude 
limit is also not entirely meaningful, as it is well known that the detection and completeness 
limits are a function of galaxy color \citep[e.g.,][]{vanderWel12}.  In order to enable 
users of the HLF GOODS-S dataset to determine the completeness limit of a given sample,  
we create an effective wavelength map that is equivalent to tracing the wavelength 
contributing the deepest data at a given location.  This effective wavelength map can 
then be used to determine the magnitude limit for any given object in the mosaic, given 
the location and $z$-$H$ color, as we describe next. 

\begin{figure}[t]
\leavevmode
\centering
\includegraphics[width=\linewidth]{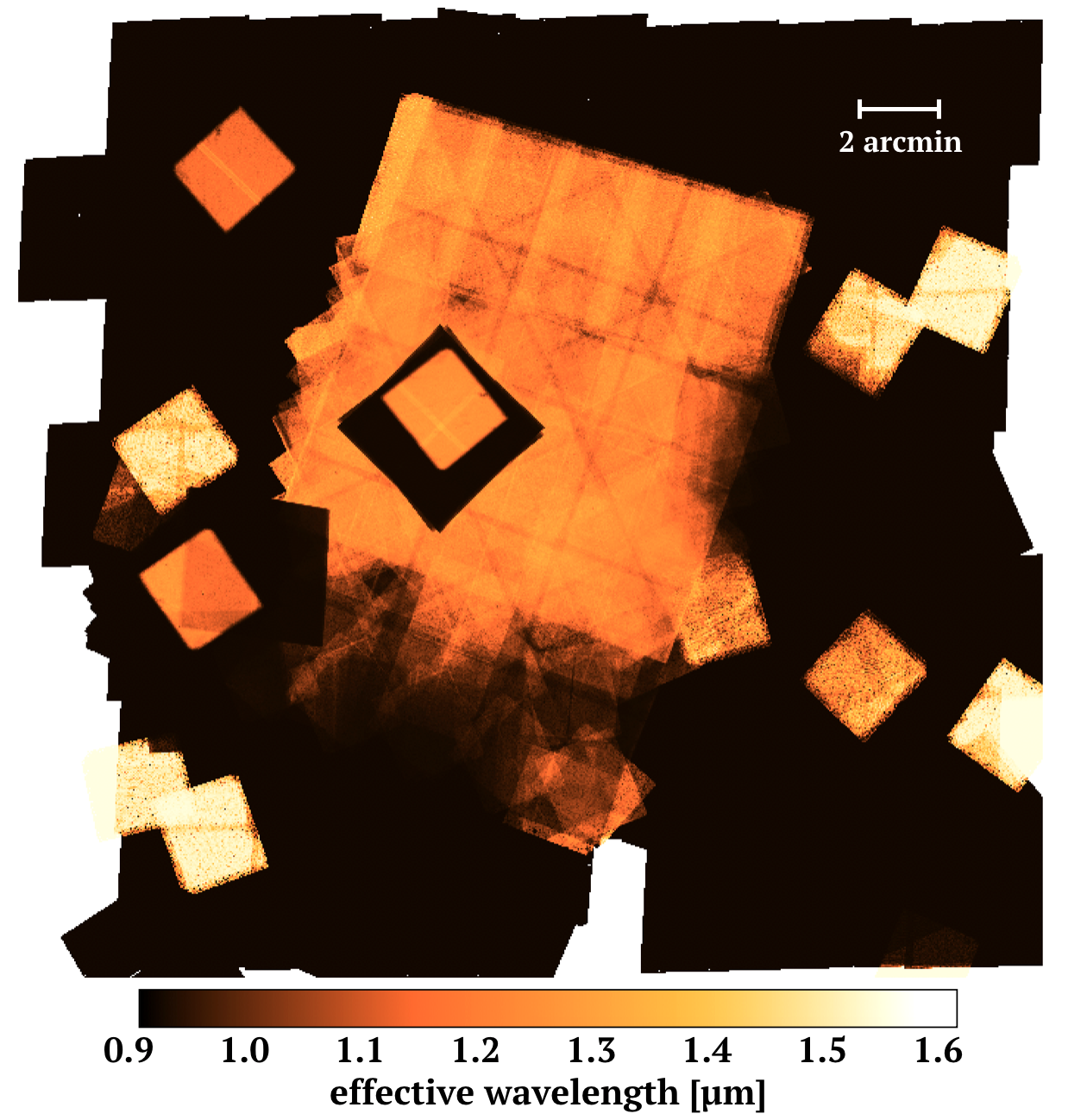}
\caption{Map of the effective wavelength of the detection image, ranging from the $z_{\mathrm{F850LP}}$ at 0.9$\mu$m (black)
to $H_{\mathrm{F160W}}$ at 1.5$\mu$m (yellow/white).  The filter with the deepest data (largest weights) will dominate the 
effective wavelength map, which varies across the field of view due to the heterogeneous nature of the HLF GOODS-S combined dataset.}

\label{fig:effwav}
\end{figure}

To create the effective wavelength map for our source detection, we take the convolved 
weight maps for each of the four filters that combine to make our noise-equalized 
detection image: $z_{\mathrm{F850LP}}$, $J_{\mathrm{F125W}}$, $JH_{\mathrm{F140W}}$, and 
$H_{\mathrm{F160W}}$.  As the released mosaics maintain the original zeropoints, we first 
correct all weight maps to a common zeropoint of 25 ABmag.  The effective wavelength map 
is then calculated as follows,

\begin{equation}
\lambda_{e} = \sqrt{\frac{\sum{(w_{X}\lambda_{e, X})^2}}{\sum{w_{X}^2}}}
\end{equation}

\noindent where $X$ corresponds to the four filters listed above, $w$ is the weight and 
$\lambda_e$ is the pivot wavelength for each filter.
Figure~\ref{fig:effwav} shows the HLF GOODS-S effective wavelength map.  The effective wavelength 
is largely representative of $\sim$1.2$\mu$m across the central field of view, with more extended 
contiguous coverage at 0.9$\mu$m.  We create an effective depth map in a similar manner as above, 
adding the four weight maps in quadrature, inverting, and taking the square root.  Given the 
effective wavelenth and depth maps, one can simply interpolate the effective wavelength at 
a given location between 0.92$\mu$m ($z_{\mathrm{F850LP}}$) and 1.54 $\mu$m ($H_{\mathrm{F160W}}$), 
using that fraction multiplied by the $z$-$H$ color to correct the effective depth in the detection 
image to the equivalent depth in the $H_{\mathrm{F160W}}$ mosaic.  In other words, 
one can approximate the effective $H_{\mathrm{F160W}}$ depth for any source from its $z$-$H$ color as:

 \begin{equation}
\sigma_{lim, H} =  \sigma_{lim, \lambda_{e}} + (z-H) \left(\frac{\lambda_{e} - 1.54}{1.54 - 0.92}\right).
\end{equation}

For example, let us consider a red object with a $z$-$H$ color of 1.0 in the mosaic outskirts where the effective wavelength 
of the detection map is 0.9$\mu$m.  If the effective depth is 26 ABmag, the depth in $H_{\mathrm{F160W}}$ would 
be 1 magnitude shallower at 25 ABmag for this source, given the red color but deeper $z$-band mosaic at this location.  
On the other hand, a blue object with a $z$-$H$ color of -1.0 in the same region would instead have an effective $H_{\mathrm{F160W}}$
depth that is 1 magnitude deeper at 27 ABmag.  
The effective wavelength and depth maps are both available to users within the larger HLF GOODS-S photometric 
catalog public release.

\subsection{Photometric Catalogs}

\subsubsection{Aperture Photometry}

Our aperture photometry methodology closely follows that of \citet{Skelton14}.  We therefore
briefly summarize the main steps followed here and note any different assumptions we have 
adopted, but defer the reader to Section 3.4 of
\citet{Skelton14} for additional details.  SExtractor is run in dual-image mode, where the 
ultra-deep noise-equalized input image is used for detection (see Section~\ref{sec:detection}) 
and the PSF-matched HST image and corresponding convolved weight map are used for the aperture photometry.  
No further background 
subtraction is needed at this stage. We perform photometry within a 0.7$^{\prime\prime}$
diameter circular aperture in all the HST bands. 
This relatively small aperture optimizes the photometry signal-to-noise ratio (SNR)
for point sources (and small higher redshift galaxies), as discussed in \citet{Whitaker11} 
and later adapted for HST resolution data in \citet{Skelton14}.  This aperture diameter was identified 
by taking a ratio of the flux enclosed from the growth curve analysis relative to the analogous 
error analysis (``empty apertures'', as described in Section 2.5.3) as a function of aperture diameter.  
The SNR peaks around 0.7$^{\prime\prime}$ for HST quality data, thus optimizing the color photometry.  
The adopted aperture corresponds to a physical radius of 2.6-3.0 kpc at $z$$\gtrsim$1, which is smaller 
than the effective radius for the majority of galaxies at these redshifts \citep{vanderWel14}.  
For the most massive galaxies, especially star-forming, the effective radii will extend beyond 
the aperture. In these cases (and at $z$$\lesssim$1), we underresolve galaxies and effectively 
measure central colors only.  The decision to adopt a relatively small aperture will therefore 
not be optimal for certain parameter spaces.  Specific examples include the majority of 
star-forming galaxies and intermediate/massive (log(M$_{\star}$/M$_{\odot}$)$>$10.5) 
quiescent galaxies at $z$$<$1, intermediate to massive star-forming galaxies (log(M$_{\star}$/M$_{\odot}$)$>$10) 
and massive quiescent galaxies (log(M$_{\star}$/M$_{\odot}$)$>$11) at 1$<$$z$$<$2, and intermediate/massive 
star-forming galaxies (log(M$_{\star}$/M$_{\odot}$)$>$10.5) at $z$$\sim$2-3.   In these cases, 
the average effective radii are similar to or larger than the adopted aperture radius due to 
their more extended light profiles.  It is worth noting that the field has not yet converged 
on the role of color gradients at high redshift.  Our methodology assumes a flat gradient 
by design, which may indeed be a fair assumption at $z$$>$2: \citet{Suess19} recently showed 
that color gradients of star-forming and quiescent galaxies are generally flat at $z$$>$2, 
but color gradients may become more prominent as redshift decreases.  In the parameter spaces 
outlined above, the spectral energy distribution (SED) will be dominated by the central light 
of the galaxy and may not be representative of the global stellar population properties.

The standard astrometry matches the CANDELS
and 3D-HST public releases, but we provide an additional column that corrects for known 
offsets in astrometry.  These astrometric differences were first detected with ALMA data, 
with offsets in the HUDF of $\delta$ra(deg)=(+0.094$\pm$0.042)/3600 and 
$\delta$dec(deg)=(-0.262$\pm$0.050)/3600 \citep[see][]{Dunlop17, Franco18}.  
We adopt an identical approach to \citet{Franco18}, but instead compare
positions in the 3D-HST photometric catalogs (which use the same astrometry as the HLF) to
the Gaia DR2 catalogs \citep{Gaia16, Gaia18}. We calculate offsets of 
$\delta$ra(deg)=(+0.011$\pm$0.08)/3600 and $\delta$dec(deg)=(-0.26$\pm$0.10)/3600, with
the equations used to define these corrections listed in Table~\ref{tab:catalog}.  
These offsets are in good agreement with \citet{Franco18}.

\begin{figure}[t]
\leavevmode
\centering
\includegraphics[width=\linewidth]{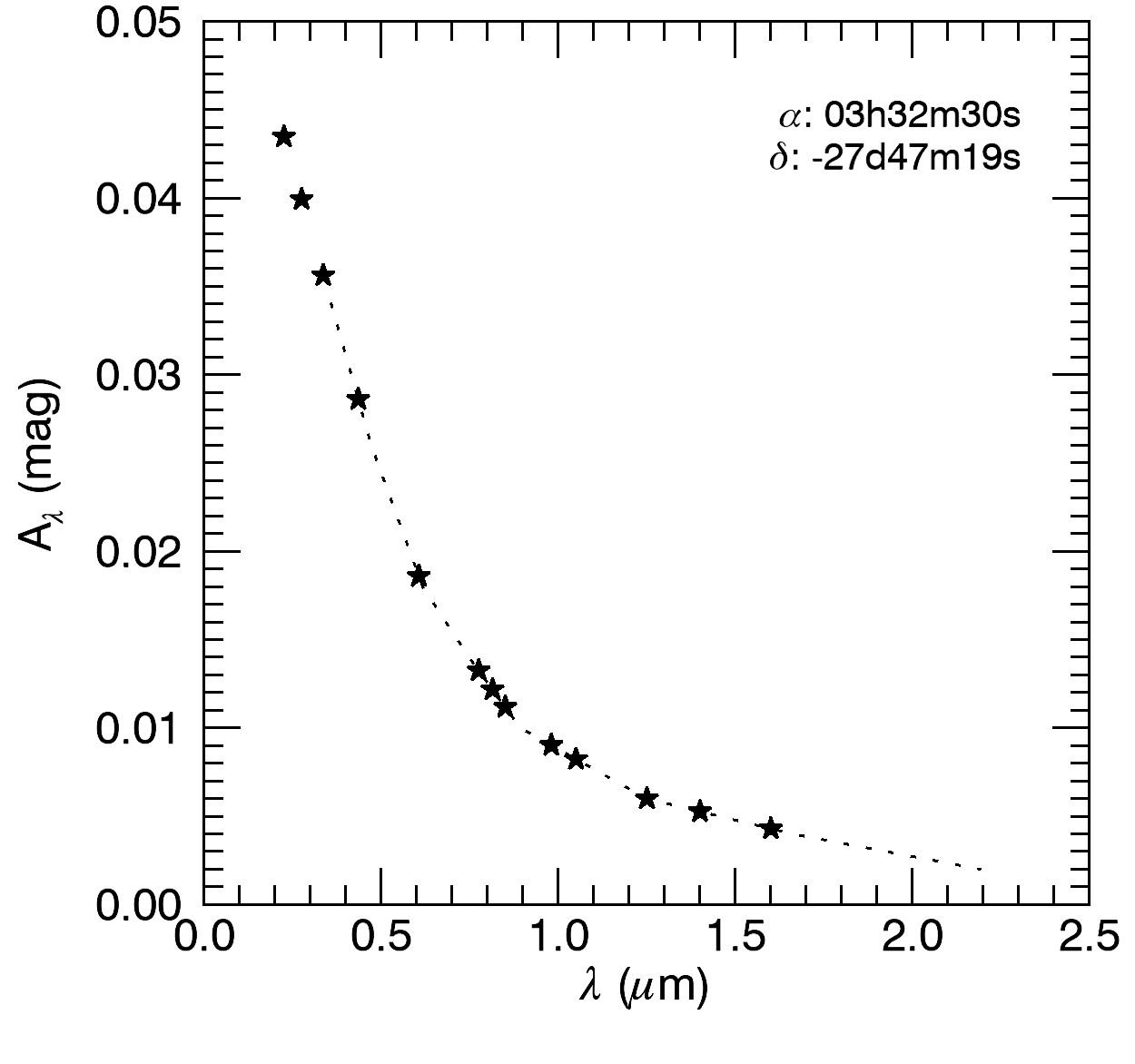}
\caption{Galactic extinction in different bandpasses from the NASA Extragalactic Database at the coordinates of the 
GOODS-S extragalactic field \citep[dotted line;][]{Schlafly11}.  Corrections for Galactic extinction are applied
to the HLF GOODS-S photometric catalog by interpolation, where the stars represent the corrections for each HST filter.}
\label{fig:extinction}
\end{figure}

The reference band is chosen to be F160W where there is coverage 
(52\% objects) and F850LP otherwise.  This decision stems from the wider area coverage of F850LP.
We return to this issue when defining columns within the photometric catalog, as there is a 
significant fraction of the mosaic with only F850LP coverage.  The total flux in the 
reference band is determined by correcting the SExtractor AUTO flux for the 
amount of light that falls outside of the AUTO aperture. 
Assuming a point source, this correction can be calculated directly from the growth curves 
described in Section~\ref{sec:PSF}. The adopted radius of the AUTO flux corresponds to the Kron 
radius\citep{Kron}, which encloses rough 90--95\% of the total light within a flexible elliptical aperture.
Our aperture correction to total flux is therefore the inverse of the fraction of light enclosed within a circular 
aperture encompassing the same area as the Kron aperture (i.e., the circularized Kron radius).
We determine this circularized Kron radius directly from the empirical growth curve for F160W
and use the same aperture correction from the reference band to scale all filters.
We apply an additional small correction ($<$0.04 mag) to the photometry to account for Galactic extinction in each
filter.  We interpolate from values given by the NASA Extragalactic Database extinction law calculator, following \citet{Skelton14} 
(see Figure~\ref{fig:extinction}). 
All fluxes within the catalog are given as total, with an AB magnitude zero point equal to 25.  
We also provide the aperture flux in the F160W and F850LP reference 
bands to allow the user to convert the total fluxes back to consistent color measurements for any band.

Unlike in \citet{Skelton14}, we do not calculate an additional photometric correction to account for any
zero point or template mismatch uncertainties.  The GOODS-S HLF photometric catalog is strictly
comprised of HST filters that typically have minimal zero point offsets calculated.  For the case of the
3D-HST GOODS-S photometric catalog, \citet{Skelton14} calculate zero point offsets ranging from 0.00 to 0.02 mag
for all filters but F435W (-0.09 mag). We will return to this point in Section~\ref{sec:compare}.

\subsubsection{Catalog Format}

The format of the photometric catalog follows that of the NEWFIRM Medium-Band Survey \citep{Whitaker11}
and the 3D-HST Survey \citep{Skelton14}, among others. The total flux and corresponding 1$\sigma$
error for each object is tabulated.  The list of column headers and their respective descriptions is located
in Table~\ref{tab:catalog}.  We briefly summarize a few notable columns below. 

The weight column for each band quantifies the relative weight for each object compared to the maximum weight 
for that filter. In practice, the weight is calculated as the ratio of
the weight at each object’s position relative to the 95th percentile of the weight map smoothed using a 
3 pixel block average. We choose to use the 95th percentile 
rather than the absolute maximum of the weight map to avoid being affected by extreme values, which is
especially important with smaller area ultra deep coverage. 
For those objects with a weight greater than the 95th percentile, we fix the value to unity in the weight column.

\begin{figure}[t]
\leavevmode
\centering
\includegraphics[width=\linewidth]{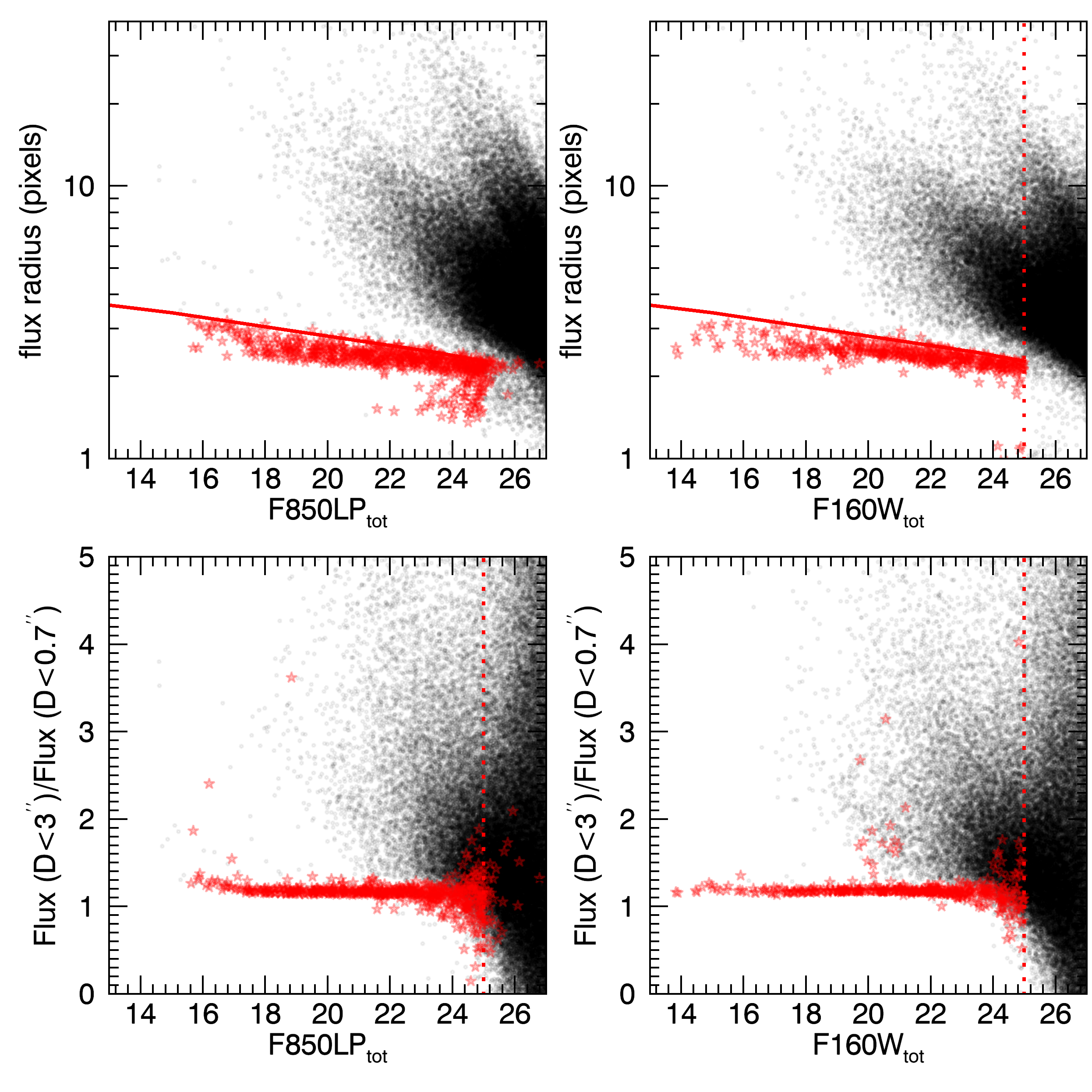}
\caption{Top panels: SExtractor’s {\tt FLUX\_RADIUS} against total $z_{\mathrm{F850LP}}$ (left) 
and $H_{\mathrm{F160W}}$ (right) magnitude. 
Objects are classified as point sources (red star symbols) in the catalog on the basis of 
flux radii less than the red line and magnitudes brighter 
than 25 ABmag (red dotted line).  Galaxies and uncertain classifications (with magnitudes 
$>$25 ABmag) are represented with black symbols. 
Bottom panels: point sources can also be classified using the ratio of fluxes in a large and 
small aperture. Although the tightness of the stellar 
sequence in this ratio at brighter magnitudes allows for a more stringent classification, 
the separation becomes less clear at fainter magnitudes. 
The flux ratio was used to select stars for the PSF-matching and kernel fitting (Figure~\ref{fig:stars}).}
\label{fig:starflag}
\end{figure}

The {\tt star\_flag} column is useful to robustly identify objects that are classified as foreground stars within 
our own Milky Way galaxy.  These point sources are identified on the basis of 
comparing their SExtractor {\tt flux\_radius} as a function of $H_{\mathrm{F160W}}$ ($z_{\mathrm{F850LP}}$) magnitude 
(Top panels of Figure~\ref{fig:starflag}). Stars 
are given a value of 1 in the {\tt star\_flag} if their flux radius falls below the selection line \citep[defined in][]{Skelton14} and 
$H_{\mathrm{F160W}}$$<$25 mag column. For
all fainter objects with $H_{\mathrm{F160W}}$$>$25 mag, we cannot robustly separate unresolved galaxies
from point sources.  These objects have {\tt star\_flag} values of 2, and we encourage the user to proceed 
with caution.  While we use the ratio in a large to small aperture as a function of magnitude 
to identify stars in the PSF-matching
section, we ultimately do not adopt this method for defining the star flag as the magnitude 
limit at which ambiguity of the tight stellar locus sets in is 
roughly two magnitudes brighter.

The {\tt detection\_flag} column has a value of unity where the F850LP mosaic is adopted as the reference
band.  In these cases, there is no F160W coverage available.  Most often, the broader wavelength coverage
in this more extended area is sparse.  In the case of no F160W coverage, all structure
parameters (e.g., {\tt kron\_radius}, {\tt a\_image}, {\tt b\_image}, {\tt flux\_radius}, etc) 
are measured from the F850LP mosaic.
Furthermore, the total fluxes are derived based on the F850LP bandpass. 

Additional noteworthy columns include the {\tt wmin\_hst} column, which indicates for any given object
the total number of HST filters with observed flux measurements.  The {\tt z\_spec} column cross-matches
the positions of each object within a radius of 0.4$^{\prime\prime}$ with the compilation of 
spectroscopic redshifts referenced in \citet{Skelton14} for the GOODS-S field.  

Finally, perhaps the two most important columns in the catalog are {\tt use\_f160w} and {\tt use\_f850lp}.  
We provide a flag within the catalog that allows a relatively straightforward selection of galaxies 
that have photometry of reasonably uniform quality. The default ``use'' flag (listed as {\tt use\_f160w} in 
the catalog, to distinguish it from spectroscopic quality flags) is set to 1 if the following criteria are met:

\begin{itemize}
 	\item Not a star, or too faint for reliable star/galaxy separation: {\tt star\_flag} = 0 or {\tt star\_flag} = 2.
	\item A detection in F160W. To limit the number of false positives, we apply a low SNR cut, requiring
{\tt f\_F160W} / {\tt e\_F160W} $>$ 3.
    \item Sufficient wavelength coverage. We require that a minimum of five filters cover the object. 
    This tends to removes objects on the
edges of the mosaics, and in gaps. When running photometric redshift or stellar population synthesis codes,
 it is common practice to require a similar threshold in the number of bandpasses.
\end{itemize}

The {\tt use\_f160w} flag selects approximately 39\% of all objects in the catalogs. Note that this 
flag is not very restrictive: for most science purposes further cuts (particularly on magnitude or SNR) 
are required. Furthermore, we caution that the flag is not 100\% successful in removing problematic SEDs. 
Generally speaking, the overall quality of an SED is higher for galaxies with a higher SNR in the WFC3 bands. 

As noted earlier, there exists wider field coverage in the GOODS-S field in the F850LP bandpass.
For this reason, we combine this filter into the ultra-deep noise-equalized detection image and adopt
it as the reference band where this is no F160W coverage.   We include the
{\tt use\_f850lp} column to indicate those objects with F850LP coverage.  The criteria used to define
this flag match the first two listed above, but for F850LP instead of F160W.  Objects with both
F160W and F850LP coverage will therefore be identified with \emph{both} use flags.  
However, {\tt use\_f850lp} is potentially more inclusive by selecting 45\% of
objects.  Though the user should be warned that not all objects selected by {\tt use\_f850lp} 
will yield robust photometric redshifts because there is no requirement
set for the minimum number of filters covered. 
To identify those objects with F850LP
coverage but no F160W coverage (47\% of objects in the catalog), the user should refer to 
the {\tt detection\_flag} column.  For these objects, the median number of HST filters with coverage is three,
enough to derive a color but not enough to derive a photometric redshift.

\begin{table*}[!th]
\centering
\caption{Catalog columns}\label{table:cat_cols}
\begin{tabular}{ll}
\hline \hline
\noalign{\smallskip}
Column name & Description \\
\hline
\noalign{\smallskip}
\texttt{id} & Unique identifier \\
\texttt{x}    &  X centroid in image coordinates\\
\texttt{y}  & Y centroid in image coordinates\\
\texttt{ra}  & RA J2000 (degrees)\\
\texttt{dec} & Dec J2000 (degrees)\\
\texttt{ra\_gaia}  & RA J2000 (degrees), corrected by Gaia astrometry following ra\_gaia(deg) = ra(deg) + 0.1130/3600 \\
\texttt{dec\_gaia} & Dec J2000 (degrees), corrected by Gaia astrometry following dec\_gaia(deg) = dec(deg) - 0.26/3600 \\
\texttt{faper\_{F160W}}    &  F160W flux within a 0.7 arcsecond aperture\\
\texttt{eaper\_{F160W}}  &1 sigma F160Werror within a 0.7 arcsecond aperture\\
\texttt{faper\_{F850LP}}    &  F850LP flux within a 0.7 arcsecond aperture\\
\texttt{eaper\_{F850LP}}  &1 sigma F850LP error within a 0.7 arcsecond aperture\\
\texttt{f\_X}   & Total flux for each filter X (zero point = 25)\\
\texttt{e\_X}  & 1 sigma error for each filter X (zero point = 25)\\
\texttt{w\_X}  & Weight relative to 95th percentile exposure within image X (see text)\\
\texttt{tot\_cor}  & Inverse fraction of light enclosed at the circularized Kron radius\\
\texttt{wmin\_hst} & Minimum weight for ACS and WFC3 bands (excluding zero exposure)\\ 
\texttt{nfilt\_hst} & Number of HST filters with non-zero weight\\ 
\texttt{z\_spec}  & Spectroscopic redshift, when available \citep[details in][]{Skelton14}\\
\texttt{star\_flag} & Point source=1, extended source=0 for objects with total $H_{F160W} \leq 25$~mag  \\
& All objects with $H_{F160W} > 25$~mag or no F160W/F850LP coverage have star\_flag = 2\\
\texttt{kron\_radius}  & SExtractor KRON\_RADIUS (pixels)\\
\texttt{a\_image} & Semi-major axis (SExtractor A\_IMAGE, pixels)\\
\texttt{b\_image}  & Semi-minor axis (SExtractor B\_IMAGE, pixels)\\
\texttt{theta\_J2000} & Position angle of the major axis (counter-clockwise, measured from East)\\ 
\texttt{class\_star} & Stellarity index (SExtractor CLASS\_STAR parameter)\\
\texttt{flux\_radius} & Circular aperture radius enclosing half the total flux  (SExtractor FLUX\_RADIUS parameter, pixels)\\
\texttt{fwhm\_image}  & FWHM from a Gaussian fit to the core (SExtractor FWHM parameter, pixels)\\
\texttt{flags}    & SExtractor extraction flags (SExtractor FLAGS parameter)\\
\texttt{detection\_flag} & A flag indicating whether the corrections and structural parameters were derived from F850LP rather than F160W \\ 
& (1 = F850LP, 0 = F160W)\\
\texttt{use\_f160w} & Flag indicating source is likely to be a galaxy with reliable measurements in $\geq$5 filters with (SNR)$_{\mathrm{F160W}}>$3 (see text) \\
\texttt{use\_f850lp} & Flag indicating source is detected with (SNR)$_{\mathrm{F850LP}}>$3 (in at least 1 filter) and likely to be a galaxy (see text) \\
\noalign{\smallskip}
\hline
\noalign{\smallskip}
\multicolumn{2}{l}{X = filter name, as defined in Section~\ref{sec:data}.}
\end{tabular}
\label{tab:catalog}
\vspace{+0.5cm}
\end{table*}

\begin{figure}[t]
\leavevmode
\centering
\includegraphics[width=\linewidth]{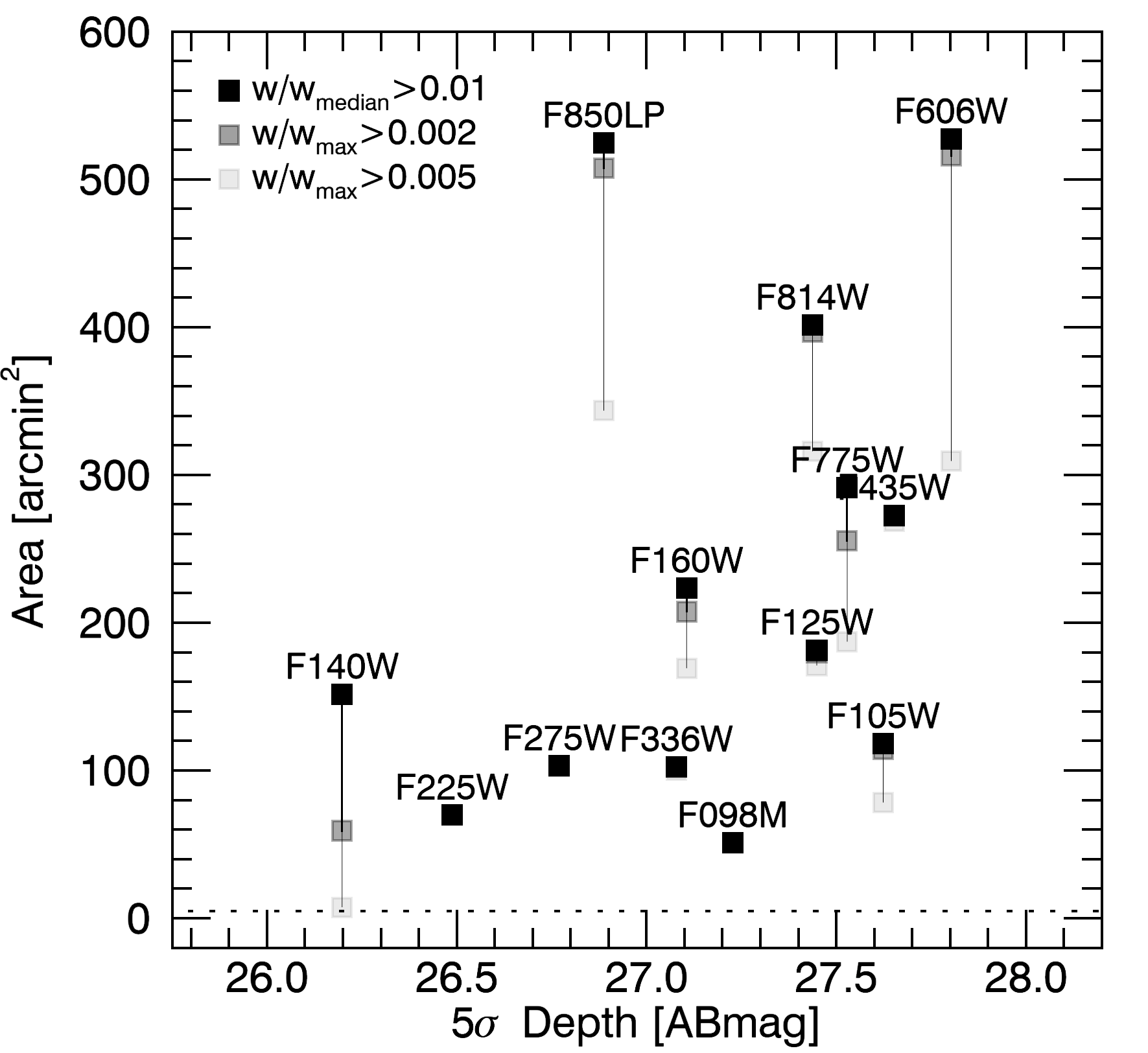}
\caption{The effective area of coverage as a function of the 5$\sigma$ point-source depths
for all 13 HST filters.  The area is calculated where the weight is greater than (1) 
1\% of the median weight (black), (2)  0.2\% of the maximum weight (grey), 
or (3) 0.5\% of the maximum weight (light grey).  The dotted line represents the area for a single
HST pointing.}
\label{fig:area_depth}
\end{figure}

\subsubsection{Error Analysis}
\label{sec:errors}

It is well known that the errors returned by SExtractor are underestimated due to the correlations 
between pixels introduced during the data reduction process. To circumvent this issue, we choose to 
measure the errors directly from the PSF-matched images themselves by placing a series of ``empty
apertures'' across the mosaics \citep[see detailed description in][]{Whitaker11}.  
Figure~\ref{fig:area_depth} shows the effective area as a function of the 5$\sigma$ point-source
depths from the empty aperture analysis.  As many of the filters include a wide range of
varying depths across the full field of view (see, e.g., Figure~\ref{fig:exptime}), we 
calculate the effective area in three different ways: we select all pixels where the weight
is greater than (1) 1\% of the median weight (black), (2)  0.2\% of the maximum weight (grey), 
or (3) 0.5\% of the maximum weight (light grey). In some filters the coverage is fairly 
homogenous (e.g., F225W-F435W, F098M, F125W), while in others there is a huge range in depth 
(e.g., F606W, F850LP, F140W).  The calculations based on the maximum weight therefore show a wide 
range of effective area for those filters that combine ultra-deep data with wide area shallower data.
For example, the vast majority of the F140W weight map 
is less than 5\% of the maximum weight, with the maximum weight originating from within 
the single UDF pointing (Figure~\ref{fig:exptime}).  
This figure therefore illustrates which filters have the most heterogenous sampling in weight, 
in addition to the typical parameter space in area and depth covered.

\begin{figure}[t]
\leavevmode
\centering
\includegraphics[width=\linewidth]{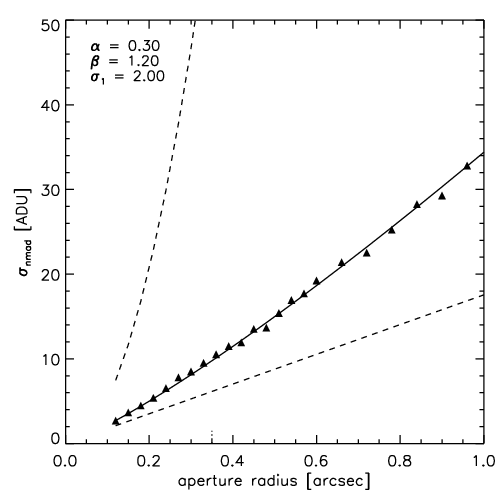}
\caption{The normalized median absolute deviation, $\sigma_{\mathrm{NMAD}}$, as a function of aperture for the F160W mosaic (triangles). 
The solid line shows the power-law fit to the data, with the best-fit parameters given in the upper left corner. 
The dashed lines indicate the case of no correlations between adjacent pixels (linear, $\propto$N) and a perfect
correlation between the pixels (N$^{2}$).}
\label{fig:kron}
\end{figure}

Given the vast range in depth across the GOODS-S field, the error analysis we adopt for the photometric
catalogs is performed on noise-equalized, PSF-matched
images. This ensures that each pixel is weighted by its corresponding depth, bringing the noise
properties to a level playing field. We measure the normalized median absolute 
deviation (nmad) from the resulting distribution of
empty aperture values for the given aperture diameter size of 0.7$^{\prime\prime}$.  
This $\sigma_{\mathrm{NMAD}}$ error is incorporated into
the catalog on an object by object basis by dividing by the square root of the weight at each
object position for each filter.  This process is repeated for a series of aperture sizes in order
to derive a corresponding error curve for the Kron radius of each individual object in the catalog 
(Figure~\ref{fig:kron}).  Given the Kron radius for any object, we use the best-fit parameters presented in 
Figure~\ref{fig:kron} to define the corresponding $\sigma_{\mathrm{NMAD}}$ error, as defined in Equation 3 of \citet{Whitaker11}.
The resulting error can be found in the \texttt{e\_X} columns within the catalog, where X corresponds to 
each respective filter.

\begin{figure}[t]
\leavevmode
\centering
\includegraphics[width=\linewidth]{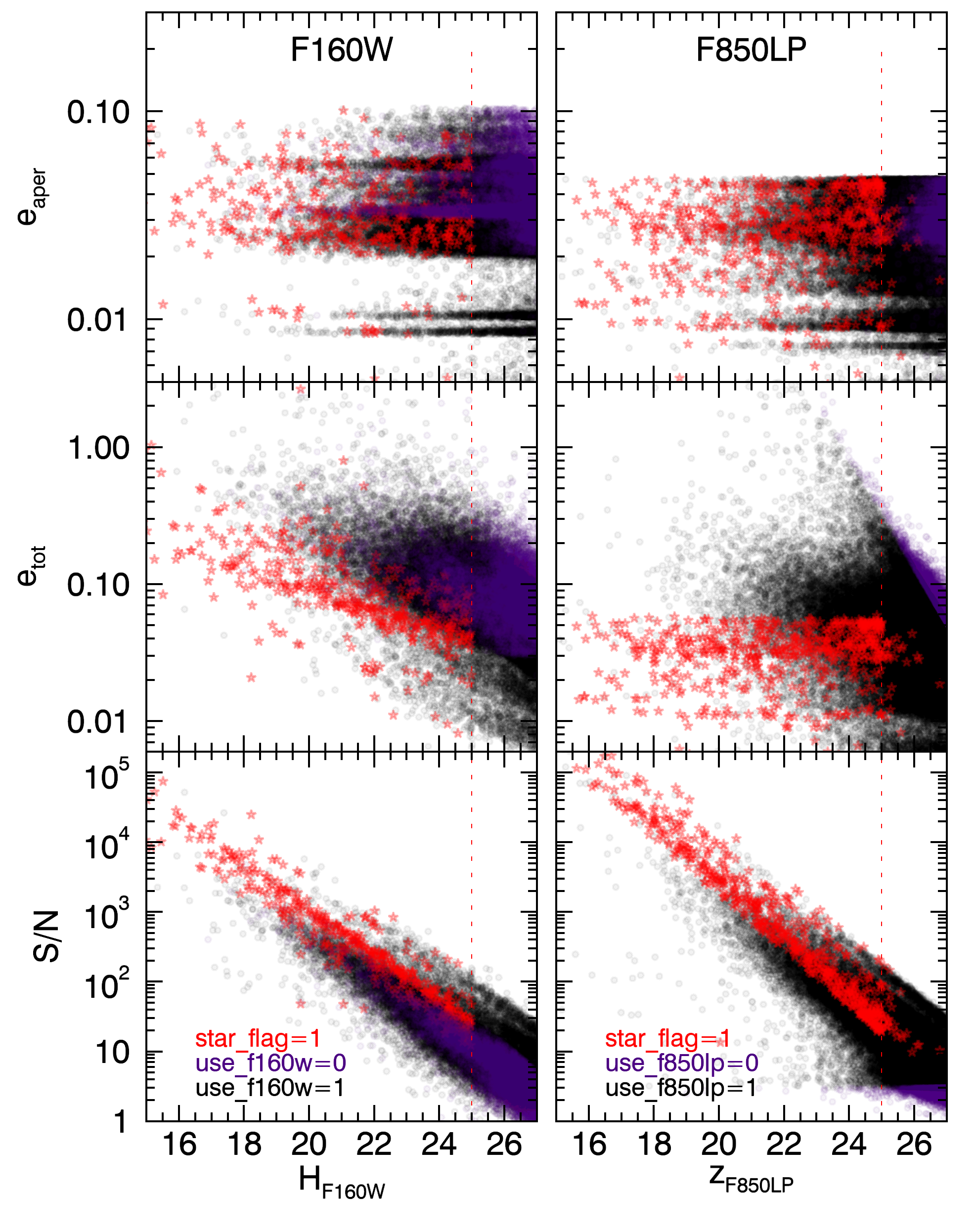}
\caption{(Top) Error as a function of $H_{\mathrm{F160W}}$ (left) and $z_{\mathrm{F850LP}}$ (right) 
within a 0.7$^{\prime\prime}$ diameter circular aperture.  Galaxies (black) are selected based on {\tt use\_f160w}=1
(left; SNR$_{\mathrm{F160W}}$$>$3, $\geq$5 HST filter coverage, not a star), compared to stars (red) and the remaining 
extended objects (purple) that do not meet this criterion (\texttt{use\_f160w}=0 and \texttt{star\_flag}$\neq$1). 
The right columns instead adopt the use\_f850lp flag, where the notable difference is that while 
the $z_{\mathrm{F850LP}}$ has wider coverage than $H_{\mathrm{F160W}}$ (343 arcmin$^{2}$ vs. 207 arcmin$^{2}$), 
most of the HST filters do not cover such a wide area. The use\_f850lp flag is therefore less restrictive, removing the 
requirement of $\geq$5 HST filters when defining use\_f850lp. (Middle) Total errors 
are scaled from the noise for the given Kron radii for each individual object, plus an extra correction to total 
based on the growth curve analysis of point sources, with the same color-coding.  (Bottom) The SNR  
generally increases with decreasing magnitude, with point sources having the highest SNRs and 
extended galaxies lower SNRs for a given magnitude.  The striping in the panels originates from the combination of 
various surveys that have a broad of depths.}
\label{fig:error_f160w}
\end{figure}

Figure~\ref{fig:error_f160w} shows the resulting $\sigma_{\mathrm{NMAD}}$ as a function of $H_{\mathrm{F160W}}$ (left panels) and 
$z_{\mathrm{F850LP}}$ magnitude (right panels), as derived from the empty
aperture methodology.  Galaxies with {\tt use\_f160w} = 1 (or {\tt use\_f850lp}) are shown in black.  
Otherwise, extended objects with {\tt use\_f160w} = 0 ({\tt use\_f850lp} = 0)
are shown in purple and point sources ({\tt star\_flag} = 1) in red. 
The top panels show the errors measured in the catalog aperture
with a diameter of 0.7$^{\prime\prime}$.  The striping is a result of combining pointings with variable
depths;  the Hubble Ultra Deep Field pointing represents the stripes with the smallest errors, whereas surveys that 
reach shallower depths but extend over wider areas will have larger errors.  The total error on $H_{\mathrm{F160W}}$ (left) 
and $z_{\mathrm{F850LP}}$ (right) is
shown in the middle panel, determined by scaling the noise (Figure~\ref{fig:kron}) to match the aperture size of the 
circularized Kron radius for each individual object and correcting to total based on the growth curve (Figure~\ref{fig:handbook}). 
More luminous objects generally have more extended light profiles, which tends to add a tilt to the total errors such 
that they scale larger at the bright end.  Finally, the lowest panels in Figure~\ref{fig:error_f160w} compare the total SNR
for $H_{\mathrm{F160W}}$ (left) and $z_{\mathrm{F850LP}}$ (right) as a function of each respective magnitude.  
Generally, point sources have the highest SNR for a given magnitude, whereas
galaxies with more extended light profiles are roughly 0.5 dex lower.  Objects with low SNR either due to intrinsic faintness 
or low weight comprise the lower envelope of the distribution of SNR vs. magnitude.  
The main difference between the {\tt use\_f160w} and {\tt use\_f850lp} flags is that the latter does not remove objects 
with less than 5 filters of coverage, resulting in a less stringent cut on the 
catalog.

\section{Data Verification}

As the 3D-HST GOODS-S photometric catalog presented in \citet{Skelton14} 
has similar F160W coverage (171 arcmin$^{2}$ vs. 207 arcmin$^{2}$) with a similar suite of bandpasses,
it serves as a natural benchmark to compare to the GOODS-S HLF photometric catalog.  In the 
following sections, we present basic comparisons between the photometry and source detection.
For all cases, we present that data when adopting either the {\tt use\_f160w} or {\tt use\_f850lp}
flags, as noted in each subsequent case. We further compare to the CANDELS GOODS-S photometric 
catalog released by \citet{Guo13}, adopting {\tt flags} equal to zero for non-contaminated sources.  
The CANDELS GOODS-S catalog presents the multiwavelength 
(UV to mid-IR) photometry, with source detection performed in the WFC3 $H_{\mathrm{F160W}}$ mosaic
using a ``hot'' and ``cold'' detection methodology.  We first present the number counts in Section~\ref{sec:numbercounts}
and then cross match all catalogs within 
a 0.5$^{\prime\prime}$ radius to compare aperture photometry in Section~\ref{sec:compare}.  
Finally, in Section~\ref{sec:sed}, we show several example SEDs to showcase the high quality of the 
photometry.

\begin{figure}
\leavevmode
\centering
\includegraphics[width=\linewidth]{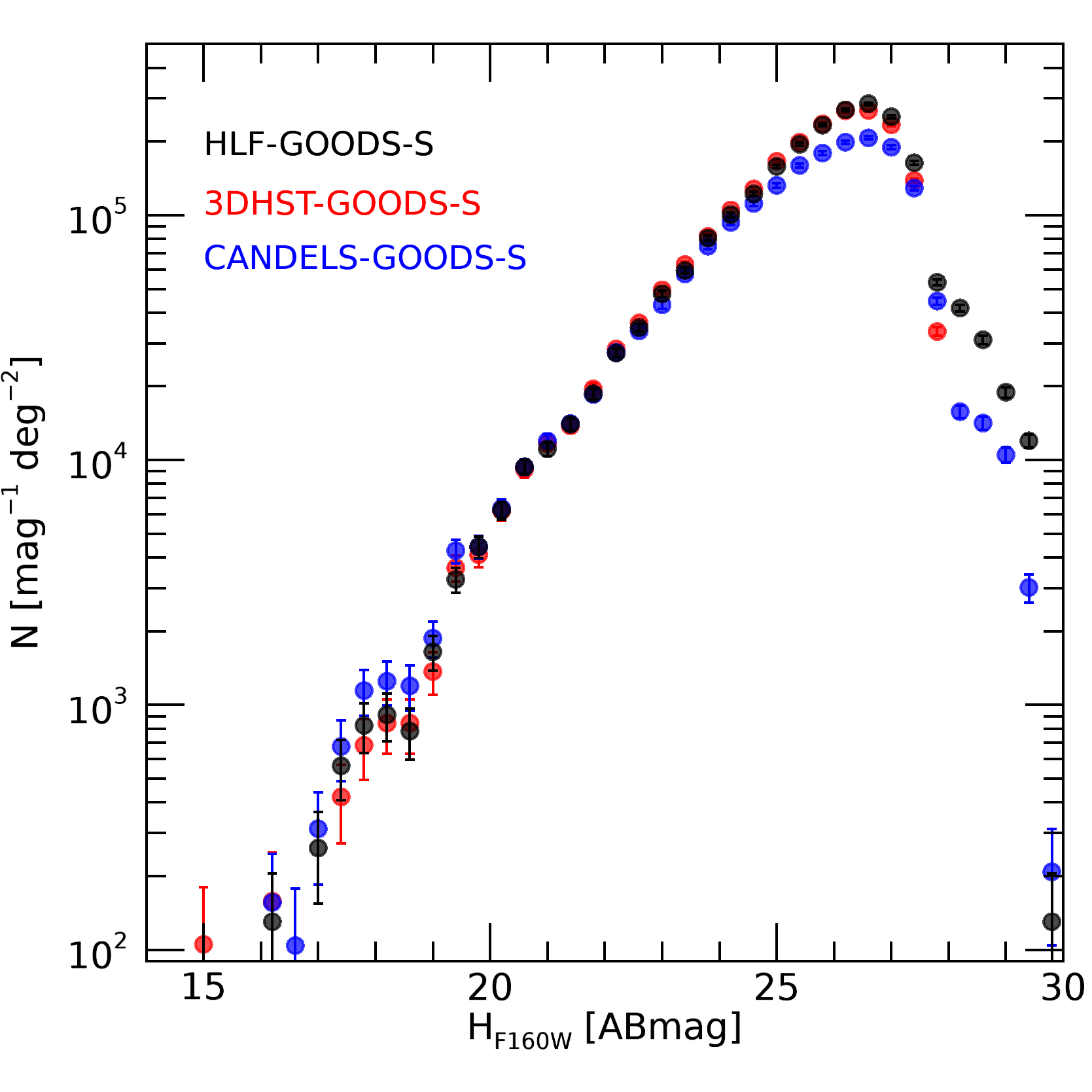}
\caption{Number counts of galaxies with Poisson errors in the GOODS-S field as a function
$H_{\mathrm{F160W}}$ total magnitude, with no correction for incompleteness.
The agreement between HLF (black), 3D-HST (red) and CANDELS (blue) is excellent.}
\label{fig:numdens}
\end{figure}

\subsection{Number Counts}
\label{sec:numbercounts}

\begin{figure*}[t]
\leavevmode
\centering
\includegraphics[width=\linewidth]{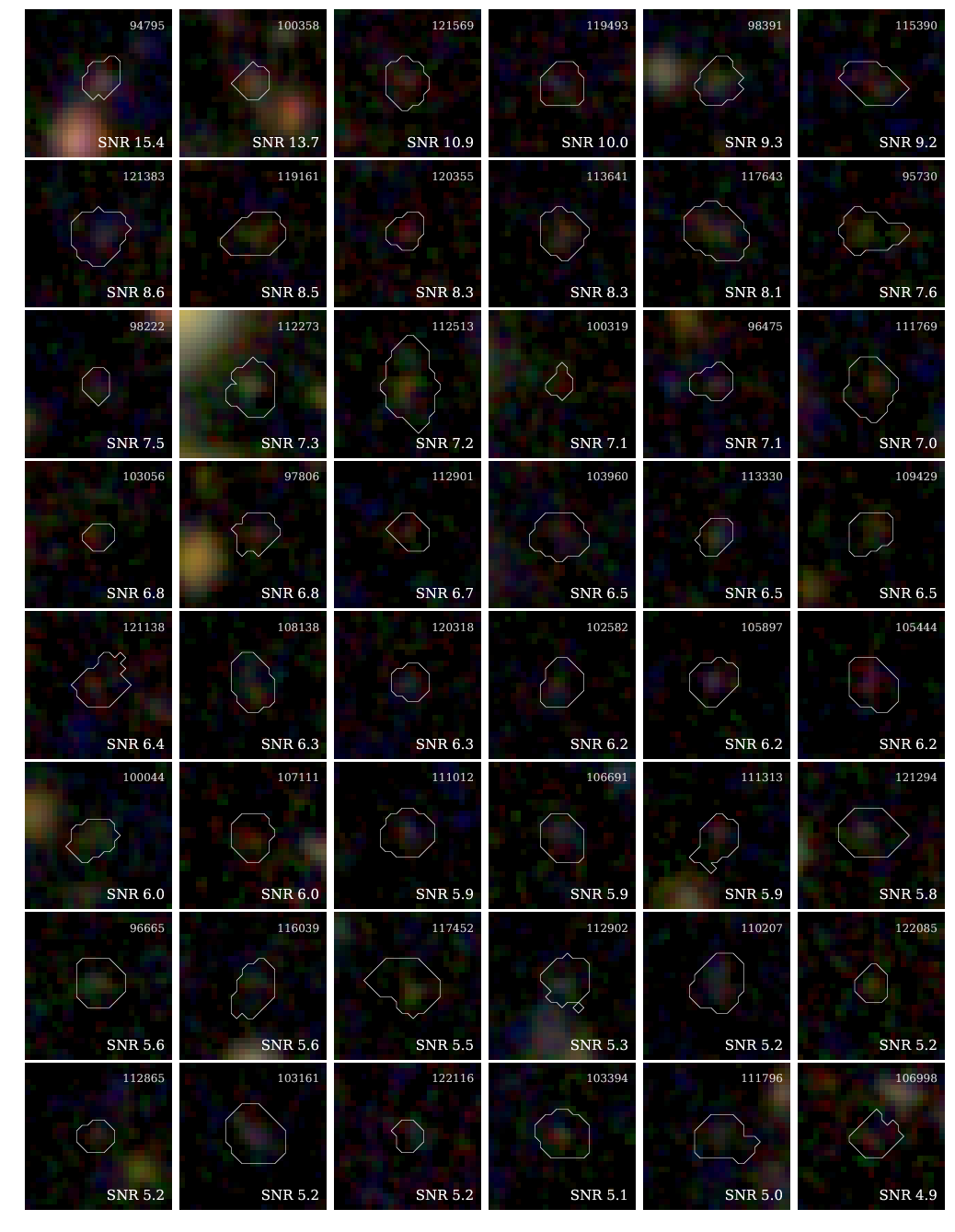}
\caption{Example postage stamps ($z_{\mathrm{F814W}}$, $J_{\mathrm{F125W}}$, $H_{\mathrm{F160W}}$) of 
48 ultra-faint sources between 28 and 29 ABmag identified in HLF but not 3D-HST.  The objects are rank ordered
by SNR, with the outline of the segmentation map shown as a white contour for reference.}
\label{fig:newpop}
\clearpage
\end{figure*}

The number density of galaxies that satisfy the {\tt use\_f160w} criterion in the GOODS-S field
are shown in Figure~\ref{fig:numdens} as a function of the total $H_{\mathrm{F160W}}$ magnitude 
for both HLF (black), 3D-HST (red), and CANDELS (blue). The error bars are Poisson. 
Though completely independent data reductions, the three data sets are fairly similar in terms of 
F160W coverage; HLF covers 207 arcmin$^{2}$ in $H_{\mathrm{F160W}}$, whereas CANDELS covers
173 arcmin$^{2}$ and 3D-HST covers 171 arcmin$^{2}$.  It is therefore
not surprising that the number counts are consistent.  The deficit of sources
with $H_{\mathrm{F160W}}$$\sim$26 ABmag in CANDELS relative to the two other fields is likely 
the result of using a deeper multi-band combined detection image. The further excess of objects at the faint end in 
HLF results from a combination of effects.  In part, this population of faint sources will 
arise due to the more aggressive source detection settings adopted.  But in some cases, it is 
clear that the HLF F160W imaging is deeper than the earlier 3D-HST version (i.e.,
explaining why both 3D-HST and HLF have more faint number counts), but HLF appears
to further reveal an exciting new population of extremely faint sources.
Figure~\ref{fig:newpop} shows 1.5$^{\prime\prime}$$\times$1.5$^{\prime\prime}$ postage 
stamps ($z_{\mathrm{F814W}}$, $J_{\mathrm{F125W}}$, $H_{\mathrm{F160W}}$) 
of 48 ultra-faint sources with magnitudes between
28 and 29 ABmag rank ordered by SNR that are identified in the HLF photometric catalog 
but do not have a match within a radius of 0.4$^{\prime\prime}$ in the 3D-HST photometric catalog. 

\begin{figure*}[t]
\leavevmode
\centering
\includegraphics[width=0.9\linewidth]{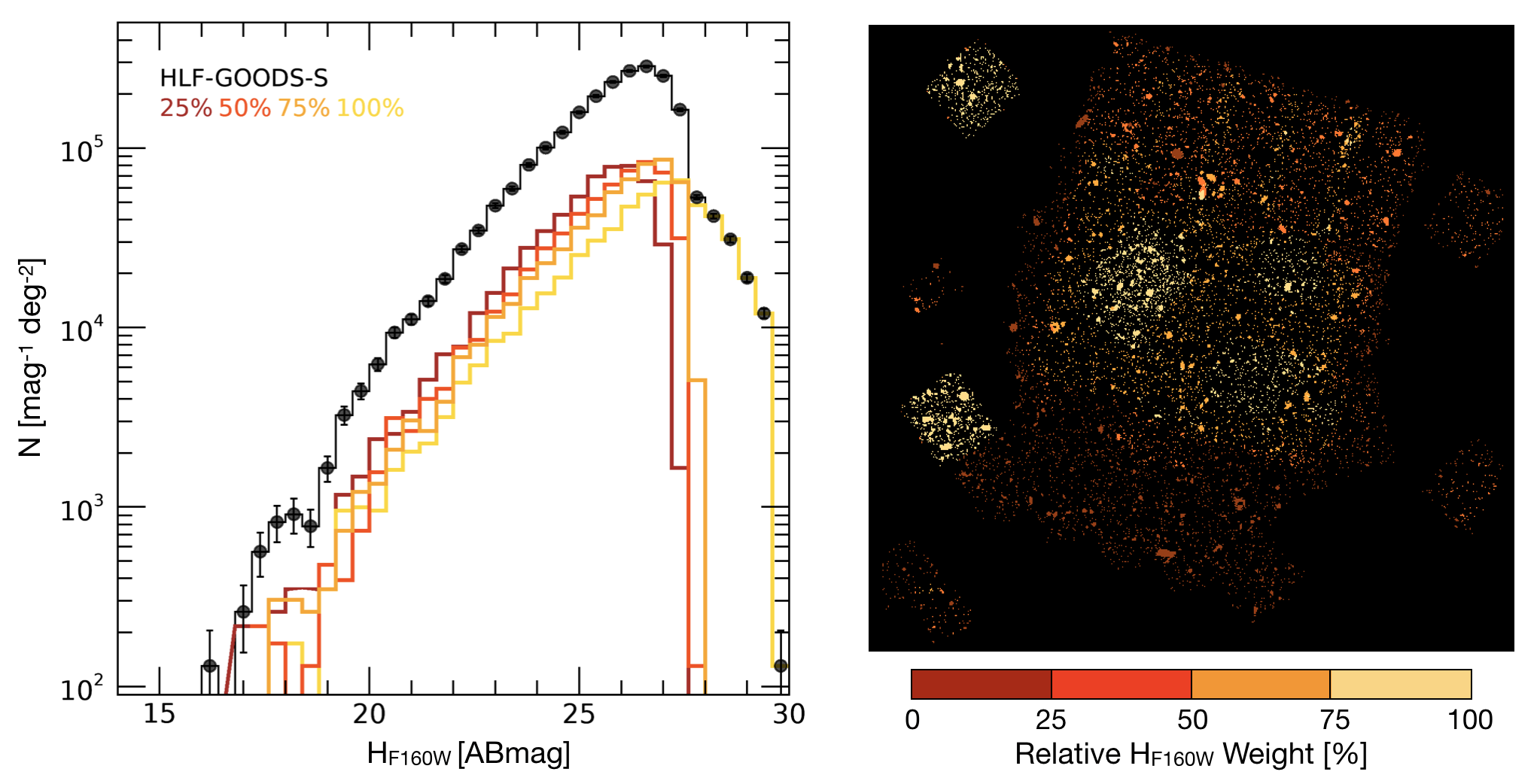}
\caption{(Left) Number counts of galaxies with Poisson errors in the HLF GOODS-S field as a function
$H_{\mathrm{F160W}}$ total magnitude, with no correction for incompleteness, broken into 
quartiles where the top quartile (white/gold) includes the deepest regions of the mosaic and the
bottom quartile (dark red) includes the shallowest coverage.  The sum of the quartiles and total
number counts is shown for reference in black. (Right) The relative weights across the segmentation map, 
color-coded by their quartile to roughly map the number counts to the on-sky location. }
\label{fig:quartiles}
\end{figure*}

The depths in the HLF GOODS-S mosaic vary significantly within the field (i.e., HUDF, CANDELS deep, wide, ERS, etc.).
Such a heterogeneous weight map implies that the single number count histogram shown in Figure~\ref{fig:numdens} 
is simply the superposition of the histograms at different depths. In order to better understand the improvement,
we separate the weight map into four quartiles that mark different depths in Figure~\ref{fig:quartiles}.
If we consider the top quartile with the highest weights (deepest data), we see that this histogram completely dominates
the faint end number counts. As expected, sources with the lowest weights (i.e., the shallowest data) are shifted towards higher magnitudes
and dominate the bright end of the number count histogram.  When combined, we recover the original distribution.
To compare the absolute and relative depths, we calculate the $H_{\mathrm{F160W}}$ depths 
in each quartile using the empty aperture method described in Section~\ref{sec:errors}.  
The HLF GOODS-S F160W mosaic reaches a 5$\sigma$ limiting point-source depth (within an aperture of radius 0.35$^{\prime\prime}$) 
of 27.0 and 29.8 ABmag in the bottom and top quartiles, respectively, with a depth of 28.7 ABmag in the middle quartiles.  
The difference between the shallow and deep regions is 3 magnitudes.  These measurements suggest that the 
HLF mosaics are deeper than the earlier compilation presented in \citet{Guo13}, given that their quoted depth 
in the HUDF is similar (29.7 ABmag), but calculated within an aperture that is a factor of two smaller.

\begin{figure}
\leavevmode
\centering
\includegraphics[width=\linewidth]{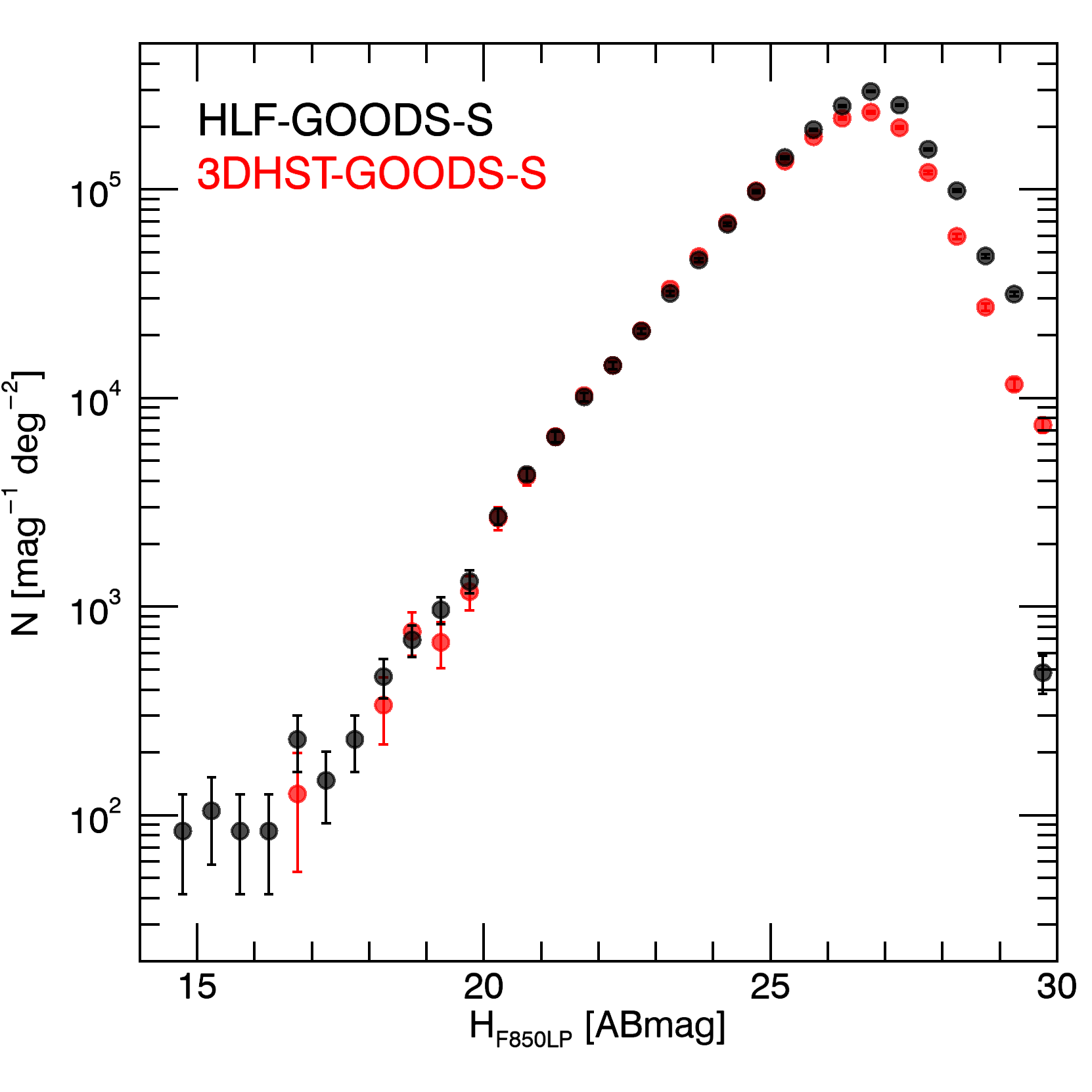}
\caption{Number counts of galaxies with Poisson errors in the GOODS-S field as a function 
$z_{\mathrm{F850LP}}$ total magnitude, with no correction for incompleteness. 
The HLF (black) covers a factor of 2 larger on-sky area (343 arcmin$^{2}$) relative
to the 3D-HST survey (red).  The agreement between the two surveys is excellent, with slight
deviations notable at the extreme bright and faint ends.}
\label{fig:numdens2}
\end{figure}

We additionally show the number density of galaxies as a function of total $z_{\mathrm{F850LP}}$ 
magnitude using the {\tt use\_f850lp} criterion in the GOODS-S field for both HLF and 3D-HST 
in Figure~\ref{fig:numdens2}. 
The total area covered within the HLF GOODS-S catalog is almost a factor of two larger than the 
3D-HST survey, with coverage for 314 arcmin$^{2}$ (assuming weights greater than 0.5\% of the maximum
weight).  Despite the significantly wider areal coverage,
the number counts reveal similar depth data when directly comparing the faint end.  However, the
advantages of surveying a wider swath of the sky is evident at the bright end, where HLF is 
able to better sample the demographics of the bright, rare galaxies. 

\subsection{Comparison with Other Surveys}
\label{sec:compare}

\begin{figure*}[t]
\leavevmode
\centering
\includegraphics[width=0.99\linewidth]{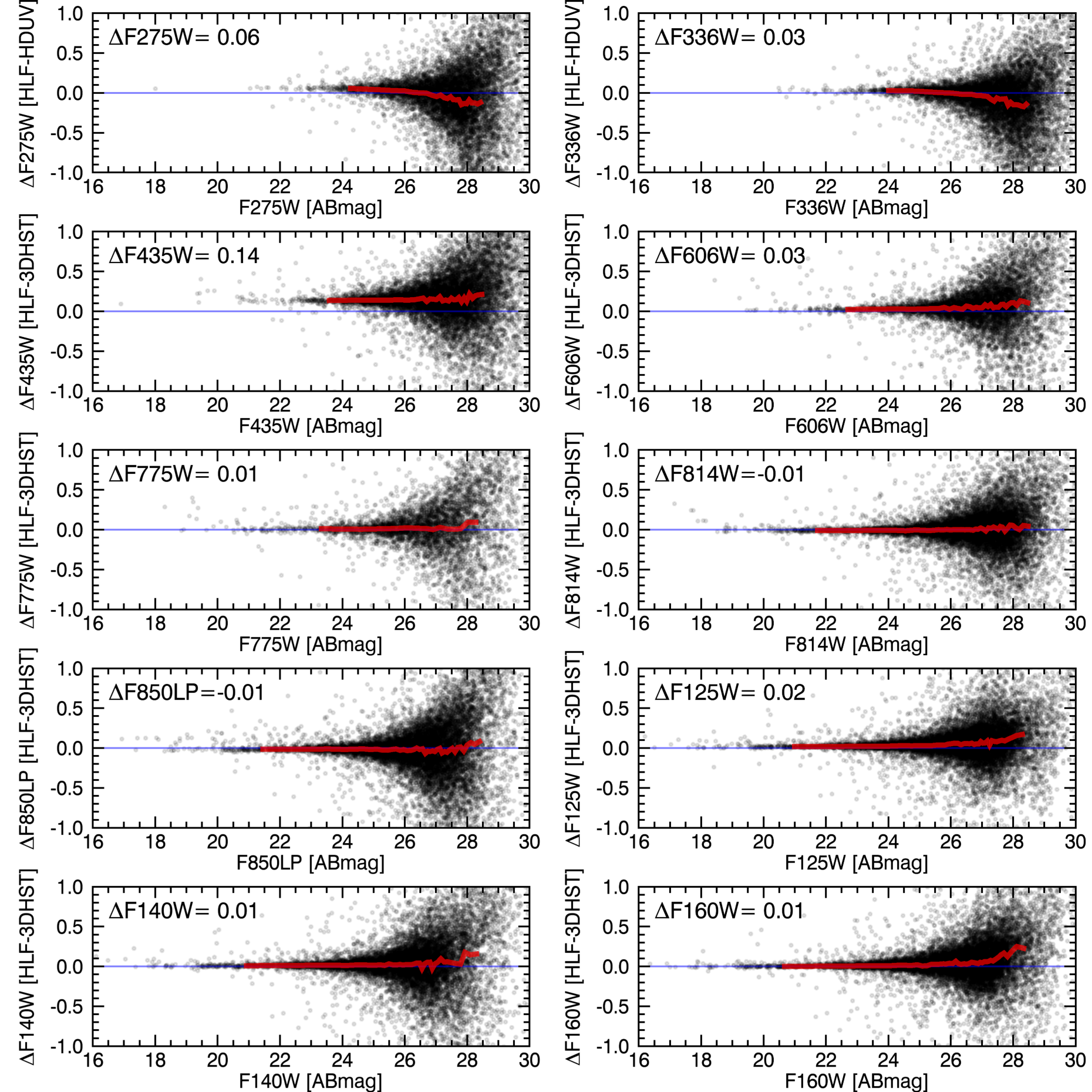}
\caption{Comparison of the GOODS-S HLF catalog to the 3D-HST \citep{Skelton14}  
and HDUV \citep{Oesch18} photometry.  We compare total fluxes from all catalogs; 3D-HST includes a zero point offset correction. 
The running median (red) line shows excellent agreement between the catalogs.
There are no significant trends with magnitude.}
\label{fig:photometry}
\end{figure*}

Measuring total fluxes for objects within any data set requires certain assumptions to be made.
It is therefore worthwhile to compare measurements to assess the quality of the photometry.
Such analyses are often invaluable in uncovering potential bugs within the catalogs.
Though the mosaics themselves were produced completely independent of one another, the 
methodology adopted to extract the photometry is largely the same between the HLF and 3D-HST
photometric catalogs.  We therefore choose to first cross-match the HLF catalog
with the v4.1.5 photometric catalogs publicly released by the 3D-HST team.  In Figure~\ref{fig:photometry}, 
we then compare the total magnitudes.  We additionally compare the HLF GOODS-S photometric catalog
to the more recent HDUV photometric catalogs for F225W, F275W,
and F336W \citep{Oesch18}, where the construction of this catalog followed the same methodology as 3D-HST and adopts
the same segmentation map.  The notable difference between the 3D-HST/HDUV and HLF catalogs is that
3D-HST performs a zero point correction, whereas we do not add this step for the HST-only HLF
photometric catalog.  The offsets between the photometry are generally quite small, 
with the red curve showing the running median in Figure~\ref{fig:photometry}.  The filter
with the largest offset is F435W.  We note that this is the same filter with the 3D-HST GOODS-S
catalog that had an offset of -0.09 magnitude applied.  When accounting for this, the 
photometry is in closer agreement relative to the original HST zero points.  Indeed,
when accounting for the zero point offset applied to the 3D-HST photometry, all HST filters
agree within $<$0.06 mag.  In other words, the photometry typically agrees at the few percent level.

\begin{figure*}[t]
\leavevmode
\centering
\includegraphics[width=0.99\linewidth]{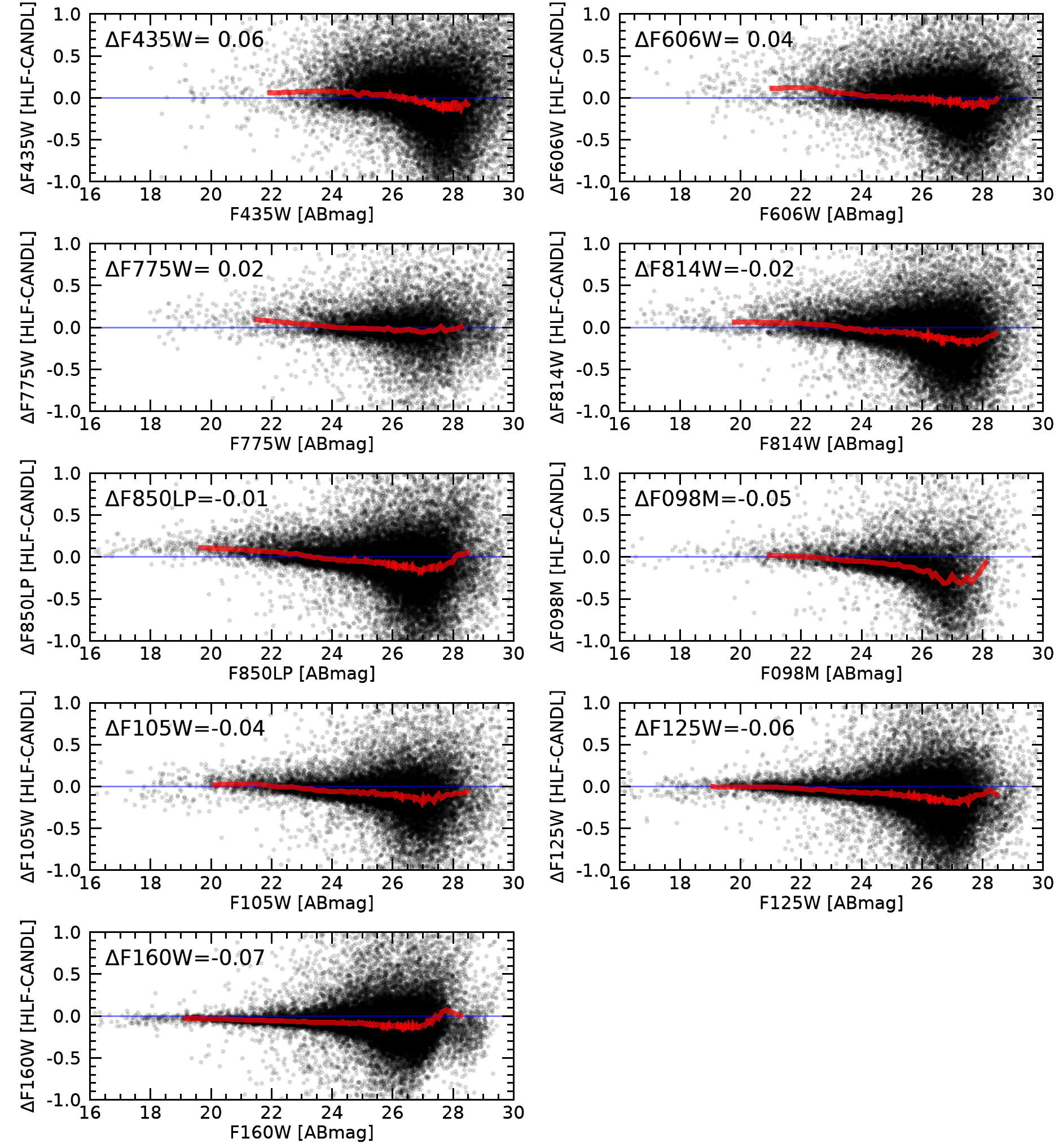}
\caption{Comparison of the GOODS-S HLF catalog to the CANDELS photometric catalog \citep{Guo13}. 
We compare total fluxes from both catalogs, where the methodologies employed by both teams are largely independent. 
The running median (red) line shows good agreement between the two catalogs.}
\label{fig:candels}
\end{figure*}

Next, we directly
compare the HST/ACS (F435W, F606W, F775W, F814W, and F850LP), and WFC3 (F098M, F105W, F125W, and F160W) total magnitudes
from the \citet{Guo13} photometric catalog to our measured photometry.
The results are shown in Figure~\ref{fig:candels}.  We find that while the analyses for the two data sets are largely 
independent of one another, the final results are consistent. There does exist a weak trend with magnitude in Figure~\ref{fig:candels}, where the CANDELS
photometry is consistently slightly fainter than HLF.  However, we note that this is only noticeable 
at the faintest magnitudes that are close to the detection limits of the data.  Overall, the two catalogs agree
remarkably well. 

\begin{figure*}[t]
\leavevmode
\centering
\includegraphics[width=0.99\linewidth]{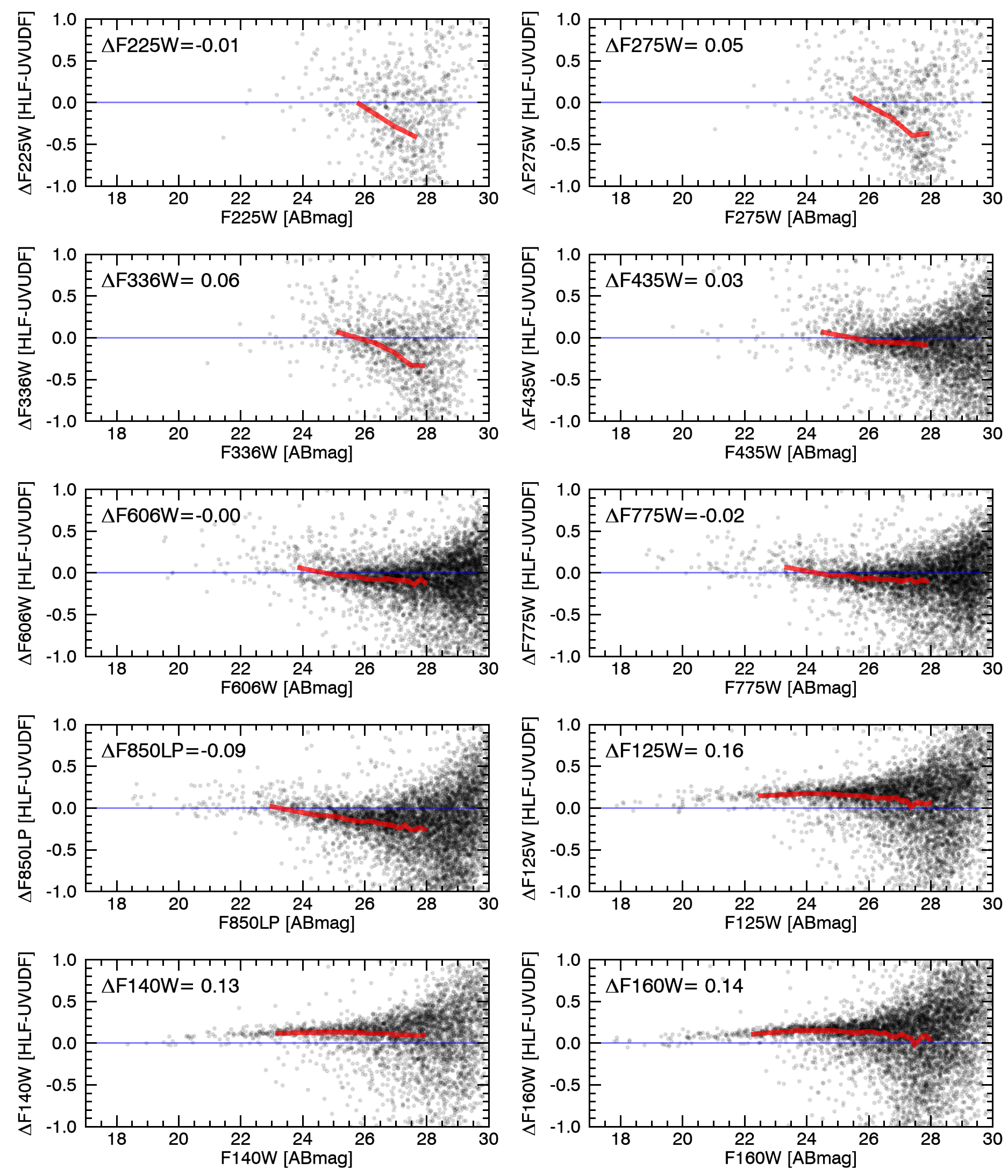}
\caption{Comparison of the GOODS-S HLF catalog to the UVUDF catalog \citep{Rafelski15}.
We compare total fluxes from both catalogs, constructed based on a different set of assumptions and algorithms. 
Overall, the running median (red) line shows good agreement between the two catalogs, with small zero point offsets at NIR wavelengths and weak
trends with magnitude in a few cases (e.g., F225W, F275W, F336W, F850LP).}
\label{fig:photometry2}
\end{figure*}

While we motivate our decisions herein for detection and analysis of mass-selected (K-band selected) samples of galaxies, 
there exist many surveys that adopt different but equally viable techniques.  
We therefore further include a comparison with UVUDF survey in Figure~\ref{fig:photometry2} \citep{Rafelski15}, 
which adopts similar methodology to the CANDELS photometric catalogs at optical and NIR wavelengths and a special analysis
of the UVIS filters.  While the UVUDF photometric catalogs measures the colors of galaxies based on 
their isophotal fluxes following
the results of \citet{Benitez04}, 
we adopt a small circular aperture flux that maximizes the SNRs.  The correction to total fluxes is also different between
the catalogs:  while both scale to total using the Kron aperture, the HLF catalog includes an additional correction 
that is typically of order 10--20\% to account for the light outside of the Kron aperture using our 
curve of growth analysis.  This explains the offset 
in the NIR filters, at least in part.  The other notable difference for the UVUDF photometry is 
that the F435W image is used as the detection when measuring the UVIS photometry, bridging between the UVIS
filters and F160W.  This results in 
slightly lower fluxes measured in the UVUDF photometry as compared to HLF, especially at the 
faintest magnitudes. Differences in background subtraction may also contribute to the discrepancies. 

\begin{figure*}[t]
\leavevmode
\centering
\includegraphics[width=0.99\linewidth]{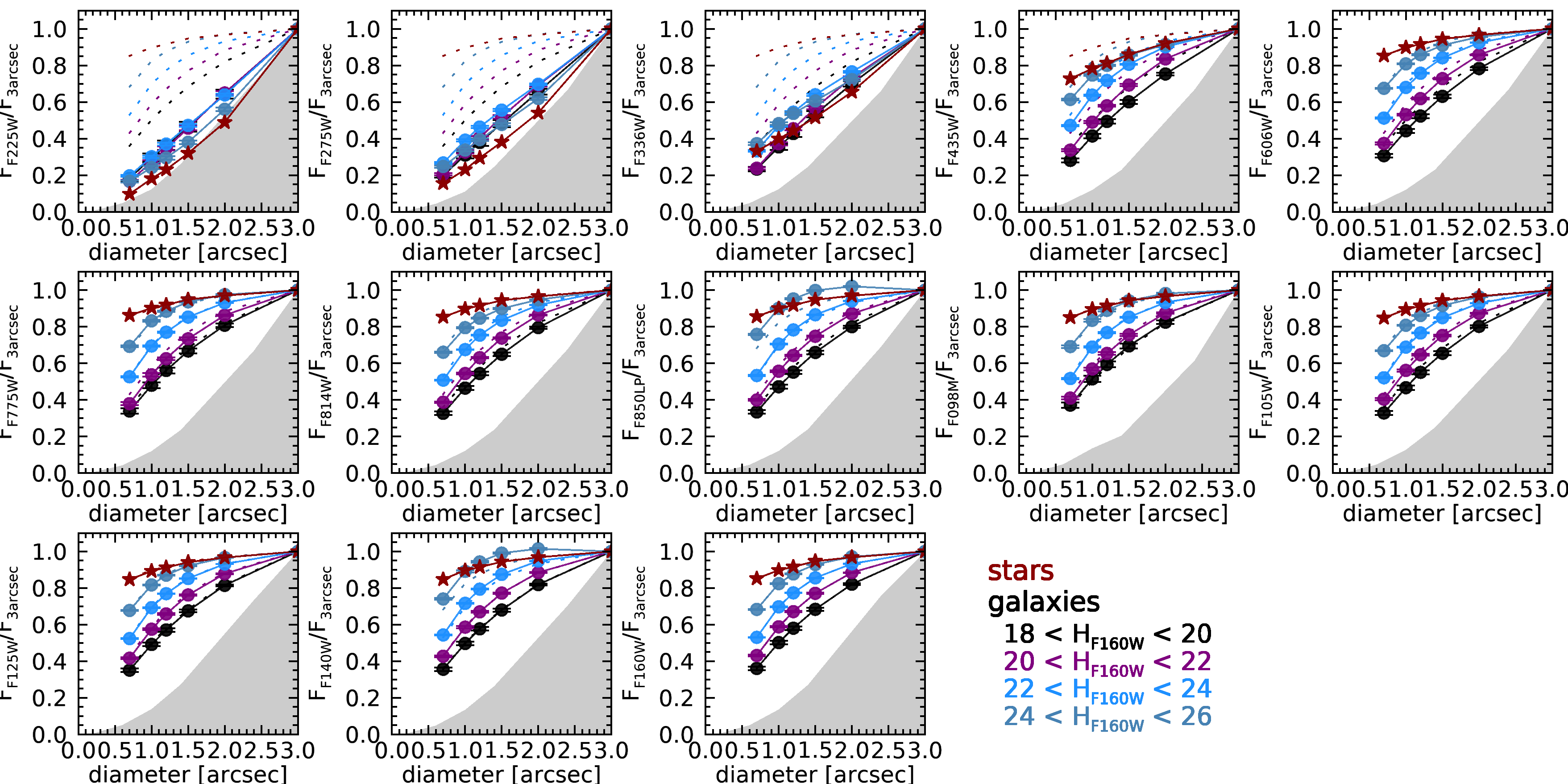}
\caption{Curves of growth for galaxies (circles, color-coded in bins of F160W magnitude) and stars, comparing
the ratio of flux in increasing aperture sizes relative to the maximum at 3 arcseconds diameter.  The
photometry is measured on the PSF-matched images and shows excellent agreement from F606W-F160W. The UVIS
filters show signification deviations due to differences in the intrinsic light profiles at these wavelengths.
The ridge of the grey shaded region is defined by the 1$\sigma$ errors derived in the empty aperture analysis.
The dashed lines are the $H_{\mathrm{F160W}}$ curves of growth, for reference.}
\label{fig:growthratio}
\end{figure*}

We explore the consequences of our IR-based detection methodology relative to the UVIS fluxes measured in 
Figure~\ref{fig:growthratio} in further detail.  Here, we select galaxies in the HLF catalog where the
\texttt{use\_f160w} flag equals unity and the SNR is greater than 20 in F160W.  
The galaxies (circles) are separated into bins of F160W magnitude
ranging from 18 to 26 ABmag, as indicated with the color-coding.  
For all galaxies within each respective bin, we measure the ratio of the flux 
within increasing circular apertures relative to a maximum aperture of diameter 3 arcseconds using 
SExtractor on the PSF matched images for the full suite of HST photometry.  The mean of this distribution is 
plotted as a function of aperture diameter, with error bars indicating the error in the mean.  We repeat this
for stars with SNR greater than 20 in F160W (red star symbols).  
For reference, we show the galaxy growth curves in the F160W image as dotted lines in all panels.
The grey shaded region demarcates the 1$\sigma$ uncertainties from the empty aperture analysis, where
any points close to this region are essentially pure noise.  
While the images used in this analysis have been homogenized, galaxies can still exhibit different intrinsic light profiles
as a function of wavelength.  This is particularly pertinent at (rest-frame) ultraviolet wavelengths, as
galaxy morphologies at these wavelengths not only can vary quite drastically outside 0.7$^{\prime\prime}$ but
their structures can also have significant differences at rest-frame optical and rest-frame FUV 
wavelengths \citep[e.g.,][]{Elmegreen07,Elmegreen09,Soto17,Guo18}.  

In Figure~\ref{fig:growthratio}, we see clear trends with magnitude that are consistent from F606W through F160W. 
Brighter galaxies are more extended and therefore have slower curves of growth, while stars have 
the most compact light profiles.  While the results are consistent in most filters, 
deviations begin to arise in the F435W filter at the 10\% level within 1 arcsecond 
and become quite dramatic in the F225W-F336W filters.  In this figure, we are comparing photometry 
for the same set of objects that have been identified and categorized based on their F160W photometry.  
The dramatic differences blueward of F435W relate to the fact that these F160W-selected objects do 
not have much intrinsic flux in the ultraviolet;  all of the magnitude bins, both bright and faint,
lie close to the 1$\sigma$ limit of pure noise (grey shaded region).  This is a known problem when trying to select stars to generate
point spread functions and hence why we identify the stars using the individual filters and not a master
list based on the deep F160W image.  

\begin{figure*}[t]
\leavevmode
\centering
\includegraphics[width=0.99\linewidth]{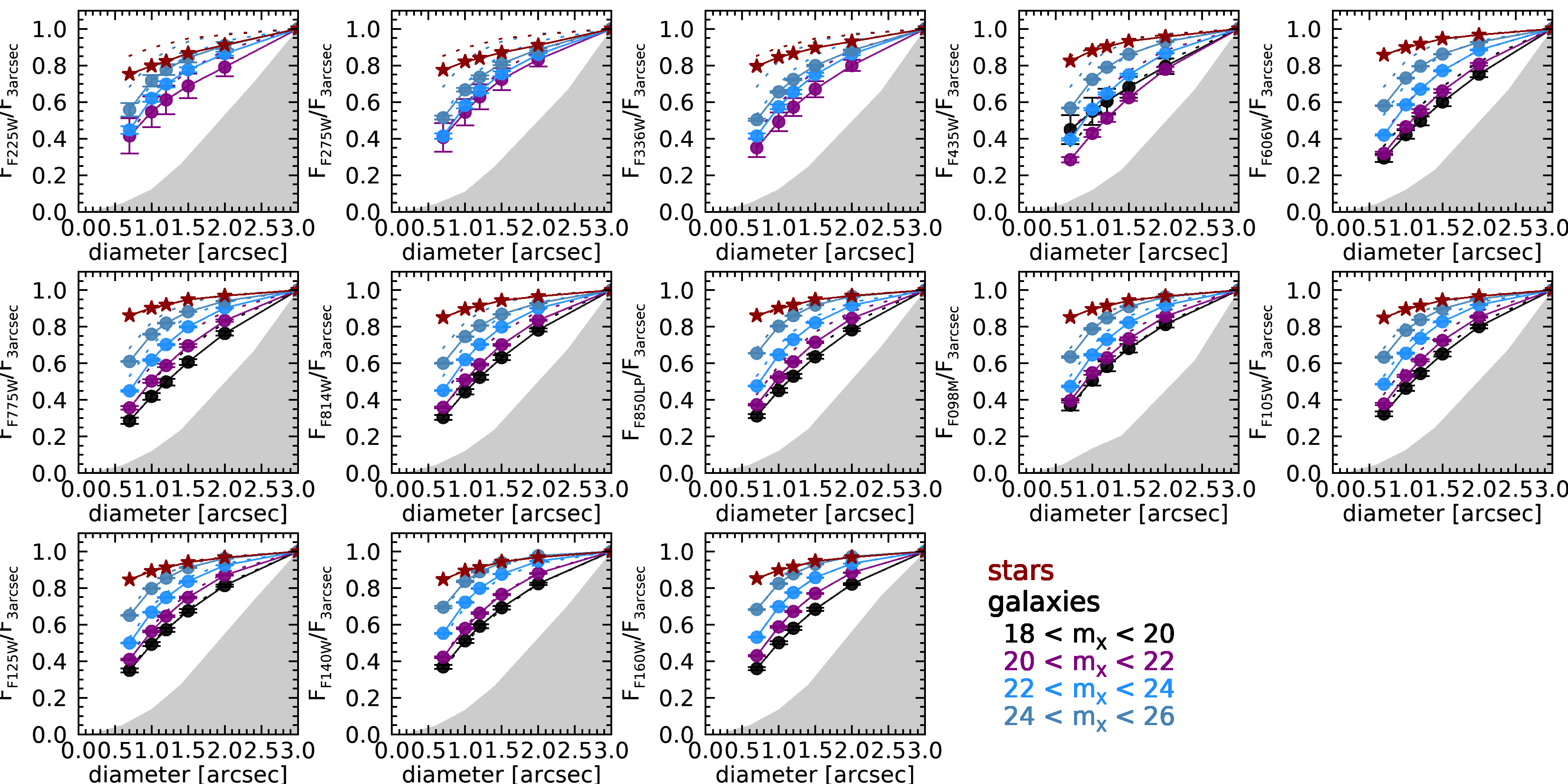}
\caption{Curves of growth for galaxies (circles, color-coded in bins of magnitude) and stars, comparing
the ratio of flux in increasing aperture sizes relative to the maximum at 3 arcseconds diameter.  With a limiting
SNR of 20, the photometry measured on the PSF-matched images shows excellent agreement from F435W-F160W, with 
slight deviations on the order of 5-10\% in the UVIS filters.   
The ridge of the grey shaded region is defined by the 1$\sigma$ errors derived in the empty aperture analysis.
The dashed lines are the $H_{\mathrm{F160W}}$ curves of growth, for reference. }
\label{fig:growthratio2}
\end{figure*}

If we instead select stars and galaxies in bins of magnitude defined separately for each filter, we 
are only considering objects that are well detected at each respective wavelength.  We compare the curves of growth
for these populations in Figure~\ref{fig:growthratio2}.  Here, we adopt the same SNR requirement of at
least 20, but in each respective filter instead of F160W alone.  This tells a slightly different story.
The light profiles based on the homogenized images are similar from F435W through F160W, with 
deviations in the UVIS filters now on the order of 5-8\% within 1 arcsecond diameter.  We suspect 
these residual differences may arise because the 
intrinsic light profiles in the UVIS filters are slightly more extended relative to the rest-optical light, 
even when convolved with the PSF.  As we correct to total flux based on the fraction of light in F160W
outside of our 0.7$^{\prime\prime}$ aperture diameter, this could result in an under-correction
at the these short wavelengths, which would serve to increase the discrepancies between the UVUDF and HLF UV photometry. 
This effect may be further augmented by the different depths in the UVIS filters; the F435W photometry 
is deeper than UVIS and also shows better agreement with the rest of the HLF photometry in Figures~\ref{fig:growthratio} 
and \ref{fig:growthratio2}.  Clumpy galaxy structure in the FUV will also contribute to the scatter,
as evident in Figure~\ref{fig:photometry2}. The main
implication of our methodology is that the SED shapes we extract are dominated by the centers of galaxies
and any strong gradients will be missed.   

\subsection{Example Spectral Energy Distributions}
\label{sec:sed}

\begin{figure*}
\leavevmode
\centering
\includegraphics[width=0.9\linewidth]{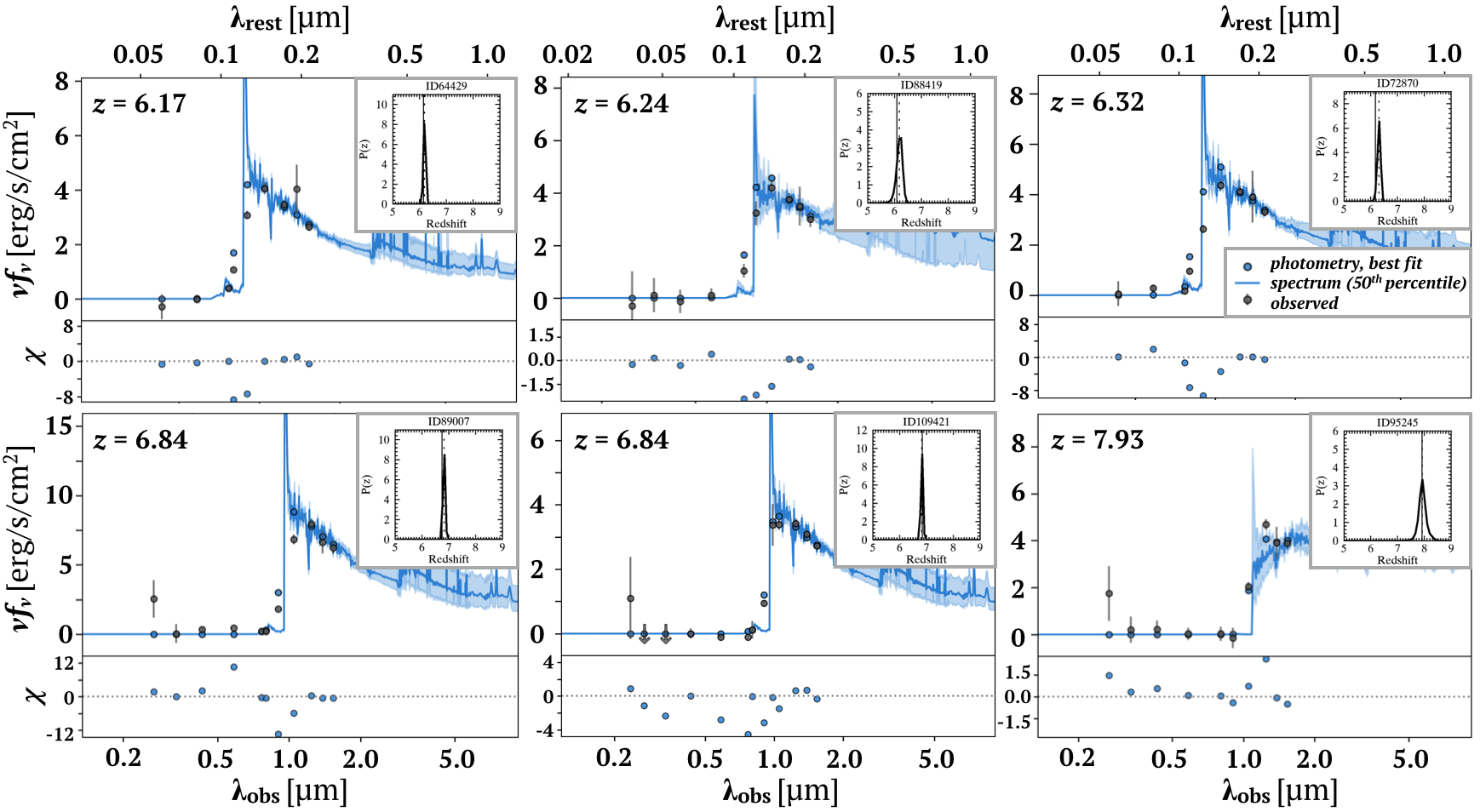}
\caption{Spectral energy distributions for six galaxies at $z$$>$6, with observed photometry (grey) and
the best-fit model (blue).  The inset panels shows the photometric redshift probability distribution for each target, 
all with single, well-defined solutions. }
\label{fig:sed}
\end{figure*}

To showcase the quality of the HLF-GOODS-S photometric catalog, Figure~\ref{fig:sed} shows the 
SEDs of a small sample of galaxies at $z$$>$6 with high SNRs in the 
near-IR HST filters.  The coverage for these 
galaxies ranges from nine to thirteen HST filters.  At these extreme high redshifts, the majority of the filters
are sampling blueward of the Lyman break.  The combination of deep, high resolution imaging with broad
wavelength coverage results in robust constraints on the photometric redshift probability distribution functions (PDF).  

Photometric redshifts are derived for these examples using the EAZY code~\citep{Brammer08}, 
which fits linear combinations of seven templates to the broadband SEDs.
This template set is optimized to be large enough to span
a broad range of galaxy colors while minimizing color and
redshift degeneracies, as described in detail in \citet{Brammer08}.
An additional template is added of an old, red galaxy,
following \citet{Whitaker11}.  We adopt \texttt{z\_peak} as the photometric redshift,
which finds discrete peaks in the redshift probability function
and returns the peak with the largest integrated probability. The inset panels
of Figure~\ref{fig:sed} show the PDFs, each with a unique, well-defined photometric redshift solution.

After fixing to the photometric redshift, we fit this high redshift sample with the \texttt{Prospector}
code, a new Bayesian framework specifically designed to use broad-band photometry to constrain high-dimensional, 
self-consistent models of galaxy formation \citep{Leja17}.  The best-fit models and realistic error bars are shown 
in Figure~\ref{fig:sed}, with stellar masses ranging from log(M$_{\star}$/M$_{\odot}$)=9.4 to 10.8.  Future forced
photometry of longer wavelength \emph{Spitzer Space Telescope} IRAC imaging will help break possible degeneracies
between dust and age, especially for the highest redshift galaxy shown here (bottom right).  
We return to the fidelity with which photometric redshifts and stellar population parameters
can be calculated based on NUV to NIR HST photometry alone to caution the users of this catalog in the following section.

\section{Summary}

In this manuscript, we describe the data analysis methodology employed to generate high
quality photometric catalogs based on the v2.0 mosaics released through the
Hubble Legacy Fields (HLF) project in the GOODS-S field.  The details of 
the data reduction can be found in \citet{Illingworth16} and \citet{Oesch18}.  

Here, we homogenize the
13 HST bandpasses, including three WFC3/UVIS filters (F225W, F275W and F336W), five ACS/WFC filters (F435W, F606W, F775W, F814W 
and F850LP) and five WFC3/IR filters (F098M, F105W, F125W, F140W and F160W). 
We use an ultra-deep detection image that combines the PSF-homogenized, noise-equalized
F850LP, F125W, F140W and F160W mosaics.  Photometry is extracted in 0.7$^{\prime\prime}$
diameter apertures and corrected to total fluxes based on the F160W curve of growth (or F850LP curve of growth in the case where there is no F160W coverage). 
The photometric catalog includes 187,464 objects, with a suggested first selection based 
either (1) {\tt use\_f160w}, which selects galaxies with SNR$>$3 in F160W and coverage in $>$5 HST bandpasses,
or (2) {\tt use\_f850lp}, which selects galaxies covering a wider on-sky area by requiring SNR$>$3 in F850LP
but no minimum coverage of HST bandpasses.

While the HLF dataset comprises the deepest mosaics of the cosmos to date, they 
are by no means meant to compensate for a lack of longer wavelength bands or more ancillary 
ground-based data.  We caution users of the HLF GOODS-S photometric catalog that deriving 
accurate stellar masses requires longer wavelength data \citep{Wuyts07,Marchesini09}. 
In particular, \citet{Muzzin09} show that including \emph{Spitzer}/IRAC data is critical 
when only broadband data (no spectroscopy) are available, improving contraints on 
M$_{\star}$, SFR, and A$_{V}$ by factors of 4, 2.5, and 0.5 mag, respectively.  
However, Muzzin et al. also show that \emph{Spitzer}/IRAC data only modestly improves the 
photometric redshifts of galaxies at z$\sim$2, whereas deep NIR photometry (such as that provided 
by the HLFs) is far more valuable in constraining photometric redshifts.  \citet{Bezanson16}
further investigate the impact of various filter combinations on the photometric redshift accuracy (see their Figure 12), 
finding that the inclusion of \emph{Spitzer}/IRAC photometry, blue (F435W) HST photometry, and 
medium-band filters particularly in the optical can have a dramatic impact \citep[see also][]{Whitaker11}. 
Relevant to the present catalog, Bezanson et al. find that the inclusion of blue (F435W) imaging in the 
3D-HST GOODS photometric catalogs significantly improves both the scatter and outlier fractions.  
As our HLF GOODS-S catalog includes additional shorter wavelenth UV data, it is relevant to note 
that \citet{Rafelski15} find similar improvements in the photometric redshifts. \citet{Rafelski15} 
demonstrate that adding NUV data to the photometric redshift derivations, in addition to the optical and NIR, 
gave a mild improvement in the scatter and a roughly a factor of 2 improvement in the outlier fraction, 
with a mild depencency on the redshift epoch under consideration.  So while the present HLF GOODS-S catalog 
will be improved in future work, with the complementary \emph{Spitzer}/IRAC analysis in particular for the 
derivation of robust stellar population physical paramaters, results in the literature confirm that 
combining HST resolution optical and NIR data with NUV already marks a notable improvement 
in the photometric redshift accuracy.

The HLF GOODS-S photometric catalog and PSF-matched mosaics and weight maps are all 
available through the HLF website (\url{https://archive.stsci.edu/prepds/hlf/}).
The HLF project and the photometric catalog presented herein will continue to serve 
the astronomical community as the next generation of space telescopes come online.  

\begin{acknowledgements}
We thank the anonymous referee for useful comments and a
careful reading of the paper.
The Hubble Legacy Fields program, supported through AR-13252 and AR-15027, is based on
observations made with the NASA/ESA Hubble Space Telescope, obtained at the
Space Telescope Science Institute, which is operated by the Association of Universities
for Research in Astronomy, Inc., under NASA contract NAS 5-26555. Financial support
for this program is gratefully acknowledged. 
The Hubble datasets used in our analysis and the generation of our catalog are from 24
programs that are listed on the MAST V2.0 HLF-GOODS-S archive.  We thank the 
numerous scientists whose programs were combined into the HLF-GOODS-S for providing a 
set of extraordinary data that will have legacy value for the community into the 
JWST era and beyond. We particularly thank Harry Teplitz, Marc Rafelski, 
Anton Koekemoer, Norman Grogin, and the UVUDF Team for providing the UVUDF dataset 
with further processing than was available publicly.
The Cosmic Dawn Center is funded by the Danish National Research Foundation.
\end{acknowledgements}


\addcontentsline{toc}{chapter}{\numberline {}{\sc References}}

\end{document}